\documentclass[aps, twocolumn,secnumarabic,nobalancelastpage,amsmath,amssymb,
nofootinbib,superscriptaddress]{revtex4-1}

\usepackage{graphics}      
\usepackage{graphicx}      
\usepackage{longtable}     
\usepackage{url}           
\usepackage{bm}            
\usepackage{xcolor}
\usepackage{mathrsfs}
\usepackage[colorlinks,allcolors=blue]{hyperref}
\usepackage{blindtext}
\usepackage{hyperref}
\usepackage{float}
\usepackage{bbold}

\begin{document}
\title{ Fast Neutrino Flavor Conversion Modes  in Multidimensional Core-collapse Supernova Models:
the Role of the Asymmetric Neutrino Distributions}

\newcommand*{\APC}{Astro-Particule et Cosmologie (APC), CNRS UMR 7164,
Universit\'e Denis Diderot, 75205 Paris Cedex 13, France}
\newcommand*{\UNM}{Department of Physics \& Astronomy, University of New
 Mexico, Albuquerque, New Mexico 87131, USA}
\newcommand*{\Numazu}{National Institute of Technology, Numazu College, Ooka 3600, Numazu, Shizuoka 410-8501, Japan}
\newcommand*{\NAOJ}{National Astronomical Observatory of Japan, Osawa, Mitaka, Tokyo 181-8588, Japan}
\author{Sajad Abbar}
\affiliation{\APC}
\affiliation{\UNM}
\author{Huaiyu Duan}
\affiliation{\UNM}
\author{Kohsuke Sumiyoshi}
\affiliation{\Numazu}
\author{Tomoya Takiwaki}
\affiliation{\NAOJ}
\author{Maria Cristina Volpe}
\affiliation{\APC}


\begin{abstract}
A dense neutrino gas, such as the one anticipated in the supernova environment, 
can experience fast neutrino flavor conversions
 on  scales much shorter than those expected in vacuum  
 probably provided that the angular distributions of $\nu_e$ and $\bar\nu_e$
cross each other.
We perform a detailed investigation of the
neutrino  angular distributions obtained by solving the Boltzmann equations for   fixed 
 matter  profiles of some representative snapshots   during the post-bounce phase of core-collapse
supernovae in 
 multidimensional  calculations 
of  an $11.2\mathrm{M}_{\odot}$ and a $27\mathrm{M}_{\odot}$ progenitor models.
Although the $11.2\mathrm{M}_{\odot}$ model features $\nu_e - \bar\nu_e$ angular crossings and 
the associated  fast modes at different
time snapshots, the  $27\mathrm{M}_{\odot}$ model
does not show any crossings 
 within the decoupling region.
 We show that  this can be understood
 by studying the multipole components of the neutrino distributions. 
 In fact, there is a higher chance for the occurrence of 
 $\nu_e - \bar\nu_e$ angular crossings for the zones where
  the multipole components of the neutrino distributions
 are strong enough. 
We also show that there can exist more than one crossings between 
the angular distributions of $\nu_e$ and $\bar{\nu}_e$.
In addition, apart from the crossings  within the neutrino decoupling region, 
there is a  class of  $\nu_e-\bar\nu_e$ angular crossings
which appears very deep inside
the proto-neutron star. 
\end{abstract}

\maketitle


\section{Introduction}
A massive star with a mass larger than $8-10$  $\mathrm{M}_{\odot}$
can end its life in a core-collapse supernova (CCSN) explosion
\cite{Colgate:1966ax}. 
During this process, a huge amount of  energy
 is released of which almost 99\% is 
in the form of neutrinos of all flavors. Although an infinitesimal fraction
of the released energy is in the form of electromagnetic radiation, only $\sim$ 0.01\%,
it is already enough to allow a CCSN to  outshine its
host galaxy.

Neutrinos can
experience flavor oscillations 
as they propagate.  
Neutrino oscillations  can, at least in principle, notably impact the physics of CCSNe.
Firstly,  it can affect the heavy elements nucleosynthesis
occurring in the SN environment by modifying
the $n/p$ ratio through the weak reactions 
  $\bar\nu_e + p \rightleftharpoons n + e^+$ and
  $\nu_e + n \rightleftharpoons p + e^-$ \cite{Qian:1996xt}. 
 Neutrino oscillations can  modify  the
$\nu_e$ and $\bar\nu_e$ spectra which in turn modifies
the rate of these reactions. 

 Moreover,  the SN dynamics can also 
 be affected by neutrino oscillations. 
 Although the mechanism of the  explosion is not well-understood yet,
 a popular mechanism proposed in
the late 1980s is the so-called \textit{delayed explosion mechanism}
in which the explosion is aided by absorbing a fraction of the energy of
 neutrinos emitted from the SN core 
\cite{Bethe:1984ux, Janka:2012wk, Burrows:2012ew}.
Likewise, neutrino oscillations can  modify  the neutrino energy deposition
by changing  $\nu_e$ and $\bar\nu_e$ spectra. 
 
Finally, to confront theory with observations, 
predictions of the neutrino fluxes and spectra are crucial for future observations 
 of CCSNe neutrino signals and the measurement of the neutrino diffuse background
\cite{Gava:2009pj, Horiuchi:2008jz, Beacom:2010kk,
   Mirizzi:2015eza,Horiuchi:2017qja, Suwa:2019svl}.

The phenomenon of neutrino oscillations in dense neutrino media,
such as the one expected in the SN environment,
 is remarkably
different from the one in vacuum and matter.    
Neutrinos
can experience collective flavor evolution
in a dense neutrino gas 
owing to the fact that 
the coherent forward scattering  
 by the background neutrino gas can now play a role in neutrino evolution
 because of the
 large neutrino number densities \cite{Pastor:2002we,duan:2006an, duan:2006jv, duan:2010bg,
  Chakraborty:2016yeg}. 
 
 Most of the initial understanding of neutrino
 flavor evolution in the SN environment  
 was based on the stationary spherically symmetric
 neutrino bulb model  \cite{duan:2006an} in which neutrinos are emitted isotropically
 from the surface of the neutrinosphere. 
  The salient feature of the results obtained 
   in this model 
  is the existence 
 of the flavor swapping phenomenon in which $\nu_e$ ($\bar\nu_e$) exchanges
 its spectra with $\nu_x$ ($\bar\nu_x$) for a range of neutrino
 energies  \cite{duan:2006jv, duan:2007sh, dasgupta:2009mg, duan:2010bg,
 Galais:2011gh, Duan:2007bt}.
  This is indeed a consequence of  collective neutrino
 oscillations in the SN environment.
 
 However, neutrinos could also undergo the so-called \textit{fast} flavor conversions
on  scales  as short as a few cm
in the densest regions of the SN core
\cite{Sawyer:2005jk, Sawyer:2015dsa,
 Chakraborty:2016lct, Izaguirre:2016gsx,  Wu:2017qpc,
  Capozzi:2017gqd, Richers:2019grc,  
  Dasgupta:2016dbv, Abbar:2017pkh, Abbar:2018beu, Capozzi:2018clo,
 Martin:2019gxb, Capozzi:2019lso, Doring:2019axc, Chakraborty:2019wxe, Johns:2019izj}. 
Unlike the traditional  collective modes which occur on scales determined
by  the neutrino vacuum frequency $\omega = \Delta m^2/2E $ 
($\sim  \mathcal O  (1)$ km for a $10$ MeV neutrino and atmospheric mass splitting),
 fast modes  occur 
 on scales $\sim G_{\mathrm{F}}^{-1} n_\nu ^{-1}$
 with $n_\nu$ and $G_{\mathrm{F}}$ being the neutrino number 
 density and the Fermi coupling constant, respectively. 
 It is believed that a necessary condition for the occurrence of fast modes is
   the presence of crossing(s) in the angular distribution of
 electron lepton number carried by neutrinos (ELN) 
 \cite{Izaguirre:2016gsx, Capozzi:2017gqd,
 Dasgupta:2016dbv, Abbar:2017pkh}. 
 It has been shown that in the presence of ELN crossings, 
 fast modes can arise due to the merging of 
 two non-collective modes \cite{Capozzi:2019lso} and  $G_{\rm{F}} n_\nu$ can play the role of $\omega$
which in turn allows for the existence of flavor conversion
modes on relatively short scales 
 \cite{Abbar:2017pkh}.

 Fast modes, if exist, 
 can remarkably  influence the physics
of CCSNe. On the one hand, they can lead to 
neutrino flavor conversions
within the SN zones that have long been thought to be the realm
of scattering processes (occurring on scales $\sim G_{\rm{F}}^{-2} E^{-2} n_{\rm{B}} ^{-1}$
with $n_{\rm{B}}$ being the baryon number density)
\cite{Cirigliano:2017hmk, Capozzi:2018clo}.
On the other hand, they 
can result in flavor conversions  close to the
surface of the proto-neutron star (PNS) where it can 
be more influential. This is important
 since (if fast modes are absent) calculations have shown so far that significant neutrino flavor conversions are not likely to occur close to the surface
of the PNS, in spite of the existence of flavor
instabilities therein
\cite{Duan:2014gfa,Chakraborty:2015tfa,Abbar:2015mca,Abbar:2015fwa, Dasgupta:2015iia}.
In fact, the unstable modes can turn stable before growing significantly due to the rapid variations of
the physical conditions during the neutrino propagation
\cite{Chakraborty:2015tfa, Hansen:2019iop}. 
However, fast modes can change this picture by 
occurring on short enough scales and therefore not being bothered 
by the rapid variations of the physical conditions.


Although such ELN crossings were not considered in the 
  bulb model calculations (neutrinos were assumed to be emitted isotropically 
 from the surface of a single sharp neutrinosphere), they are speculated  
 to exist in realistic SN models. This simply arises from
 the fact that 
 $\nu_e$ and $\bar\nu_e$ decouple at different radii. Thus, one might simply 
 expect that ELN crossings should be unavoidable 
 in the SN environment \cite{Sawyer:2015dsa, Chakraborty:2016lct}.

Nevertheless,
one-dimensional SN simulations have not shown  such ELN crossings
during the early stages of  CCSNe within the 
shock region \cite{Tamborra:2017ubu, Shalgar:2019kzy} (note,
however, that  they 
might still exist in the pre-shock SN region \cite{Morinaga:2019wsv}).
 Although the angular distributions of $\bar\nu_e$'s are
 normally more peaked  than
 that of $\nu_e$'s in the forward direction, the large difference between
 the number densities of $\nu_e$ and $\bar{\nu}_e$
 hinders the occurrence of ELN crossings.
 However, this story can be changed in multidimensional (multi-D)
 SN models. Indeed, recent multi-D SN
 simulations have shown that the
 neutrino distributions can be highly asymmetric  in the 
 presence  of lepton-emission self-sustained asymmetry (LESA) 
 \cite{Tamborra:2014aua, Vartanyan:2019ssu, Sugiura:2019xuv, Walk:2019ier, Glas:2018vcs,
 OConnor:2018tuw, Janka:2016fox, Tamborra:2014hga, Nagakura:2019evv}.
 Such  asymmetric neutrino distributions   can significantly    help increasing the chance of 
 the occurrence of ELN crossings by providing SN zones
 with smaller difference  between $n_{\bar\nu_e}$
 and $n_{\nu_e}$ \cite{Abbar:2018shq}.
 
 However, most of the state-of-the-art multi-D SN 
simulations do not provide such detailed angular information of
 neutrinos  due to
the simplifications made in the neutrino transport. 
SN simulations in which full  neutrino
 angular distributions are available have  become accessible
just  recently \cite{Sumiyoshi:2012za, Sumiyoshi:2014qua, Nagakura:2017mnp}. 
The first investigation of the occurrence of ELN crossings in multi-D SN models was reported in Ref.~\cite{Abbar:2018shq} 
in which the Boltzmann equations were solved for a number of
fixed SN matter profiles and 
a number of ELN crossings and the associated fast modes were found in both 2D and 3D calculations of an
$11.2\mathrm{M}_{\odot}$ progenitor model.
Subsequently, the results of a self-consistent 2D
SN simulation (solving the Boltzmann  equations and hydrodynamics simultaneously) were reported 
in which ELN crossings were not observed for the selected few spatial points \cite{Azari:2019jvr}.
Although both of the  computations were performed for an 
$11.2\mathrm{M}_{\odot}$ progenitor model, the employed equations of state (EOS)
were different. This can have profound consequences for
the occurrence of ELN crossings, as will be discussed in this paper.

In this study, we explore 
the neutrino angular distributions 
obtained by solving 
 the Boltzmann  equations for several fixed SN matter profiles 
  which are representative snapshots 
 taken from  multi-D SN simulations.
Following our previous study \cite{Abbar:2018shq}, 
we present a more detailed investigation of
the ELN crossings in an $11.2\mathrm{M}_{\odot}$ progenitor  model.
Furthermore, we present and analyse our results of the calculations of 
 a $27\mathrm{M}_{\odot}$ progenitor models
 in which no ELN crossings within/above the neutrino decoupling
 region  were observed (Sec.~\ref{sec:ELNcrossings}). 
We also discuss the possibility of the
existence of a class of ELN crossings in very deep regions
 well inside the PNS 
 (Sec.~\ref{sec:PNS})\footnote{Refs.~\cite{Nagakura:2019sig, DelfanAzari:2019tez} 
 appeared while this manuscript was in the last stages of its preparation.
In Ref.~\cite{DelfanAzari:2019tez}, the authors  report similar  occurrence of  deep ELN crossings
inside the PNS.}.
By performing linear stability  analysis (Sec.~\ref{sec:linear}), 
we show that  fast modes  associated with the ELN crossings
can lead to flavor conversion rates as large as  several 
e-folds per nanosecond (Sec.~\ref{sec:ELNcrossings}).


\section{Linear stability analysis.---} \label{sec:linear}

 The state of a neutrino traveling with momentum $\mathbf{p}$ 
  can be specified by its flavor density matrix $\varrho_{\mathbf{p}}(t, \bf{x})$ \cite{Sigl:1992fn}
  at each time $t$ and point $\textbf{x}$, 
and its flavor evolution in the absence of the collision term is governed by the
Liouville-Von Neumann mean-field equation of motion 
\cite{Sigl:1992fn,Strack:2005ux,Cardall:2007zw,Volpe:2013jgr, Vlasenko:2013fja}
\begin{equation}
i (\partial_t + \mathbf{v} \cdot \nabla) \varrho_{\mathbf{p}} = [\rm{H}_{\mathbf{p}},\varrho_{\mathbf{p}}],
\label{Eq:EOM}
\end{equation}
with $ \mathbf{v} $ being the neutrino velocity
which in the spherical coordinate can be defined as  $ \mathbf{v} = 
(\sin\theta_\nu \cos\phi_\nu, \sin\theta_\nu \sin\phi_\nu, \cos\theta_\nu)$
 for a neutrino with emission angles ($\theta_\nu$, $\phi_\nu$).
 Also, $\mathrm{H}_{\mathbf{p}} = \mathrm{H_{vac}} + \mathrm{H_{mat}} + \mathrm{H}_{\nu \nu, \mathbf{p}}$ 
 is the total  Hamiltonian
where, in the two-flavour scenario, 
\begin{align}
\rm{H_{vac}} &\approx \frac{\eta\omega}{2}
\left[ {\begin{array}{cc}
1   &   0  \\
0  & -1 \\
\end{array} } \right] \\
 \rm{H_{mat}}  &= \frac{\lambda}{2}
\left[ {\begin{array}{cc}
1   &   0  \\
0  & -1 \\
\end{array} } \right] 
\end{align}
with $\eta = +1  (-1)$ for the inverted (normal) mass ordering
 and $\lambda = \sqrt2 G_{\mathrm{F}} n_e$ with $n_e$ being the electron number density \cite{Wolfenstein:1977ue,Mikheev:1986gs}, 
where it is assumed that $\theta_{\rm{v}} \ll 1 $
and strong matter currents are absent. 
Finally,
\begin{equation}
\begin{split}
 \mathrm{H}_{\nu \nu, \mathbf{p}} = \sqrt2 G_{\rm{F}}  \int  \frac{\mathrm{d}^3\mathbf{p'}}{(2\pi)^3} 
  \big( 1- \mathbf{v} \cdot \mathbf{v'}  \big)
   \big( 
 \varrho_{\mathbf{p'}}
 -   
 \bar{\varrho}_{\mathbf{p'}} \big),
   \end{split}
 \end{equation}
 is the contribution from neutrino-neutrino refraction \cite{Fuller:1987aa,Notzold:1988kx,Pantaleone:1992xh}.
 Here, the flavor density matrices can be written as \cite{Banerjee:2011fj}
 \begin{align}
  \varrho = \frac{f_{\nu_e} + f_{\nu_x}}{2} \mathbb{1}
  + \frac{f_{\nu_e} - f_{\nu_x}}{2} \begin{bmatrix}
    s & S \\ S^* & -s \end{bmatrix},
\end{align}
where $S$ and $s$  are some complex and real quantities, respectively,
and  $f_{\nu}$ is the (prior to oscillation)  neutrino occupation numbers, 
so that  the neutrino number  and flux densities are
 \begin{equation}
\begin{split}
n_\nu &= \int\!\frac{\mathrm{d}^3 p}{(2\pi)^3} f_\nu(\mathbf{p}),\\
\mathbf{j}_\nu &= \int\!\frac{\mathrm{d}^3 p}{(2\pi)^3}
f_\nu(\mathbf{p}) \mathbf{v},\\
\end{split}
 \end{equation}
respectively. Likewise, $\bar\nu$ quantities can be defined with respect to 
$f_{\bar\nu_e}$ and $f_{\bar\nu_x}$.

 In the presence of fast modes, Eq.~\eqref{Eq:EOM} turns out to be approximately blind to 
the neutrino energy (at least in the linear regime).
 Thus, one
can set $\omega=0$ and drop the energy dependency of $\varrho$
and integrate over the neutrino energy. It then proves to be useful 
to define the neutrino electron lepton number (ELN) as  \cite{Izaguirre:2016gsx}
\begin{equation}
  G_\mathbf{v} =
  \sqrt2 G_{\mathrm{F}}
  \int_0^\infty \frac{E_\nu^2 \mathrm{d} E_\nu}{(2\pi)^3}
        [f_{\nu_e}(\mathbf{p})- f_{\bar\nu_e}(\mathbf{p})],
 \label{Eq:G}
\end{equation}
assuming $f_{\nu_x}(\mathbf{p}) = f_{\bar{\nu}_x}(\mathbf{p})$.

One can then linearise Eq.~\eqref{Eq:EOM} by keeping the 
terms of order $\mathcal{O}(|S_\mathbf{v}|)$ or larger
\cite{Banerjee:2011fj, Vaananen:2013qja, Izaguirre:2016gsx},
\begin{equation}
 i (\partial_t + \mathbf{v} \cdot \bm{\nabla}) S_\mathbf{v}
 = (\epsilon_0 + \mathbf{v} \cdot
 \boldsymbol{\epsilon} ) S_\mathbf{v}
 - \int\!\mathrm{d}\Gamma_{\mathbf{v}'} (1 - \mathbf{v} \cdot \mathbf{v}')
 G_{\mathbf{v}'} S_{\mathbf{v}'},
 \label{Eq:linear}
 \end{equation}
 where $\mathrm{d}\Gamma_{\mathbf{v}'}$ is the differential solid angle in the
 direction of $\mathbf{v}'$,
 $\epsilon_0 = \lambda+\int\!\mathrm{d}\Gamma_{\mathbf{v}'} G_{\mathbf{v}'}$,
 and $\bm{\epsilon} = \int\!\mathrm{d}\Gamma_{\mathbf{v}'}
 G_{\mathbf{v}'} \mathbf{v}'$.

Because Eq.~\eqref{Eq:linear} is linear in $S_\mathbf{v}$,
it has the normal solutions of the form 
\begin{equation}
S_\mathbf{v}  = Q_\mathbf{v}
 e^{ -i\Omega t  + i{\textbf{K}} \cdot {\textbf{x}}}.
 \label{Eq:exp}
 \end{equation}
 Collective neutrino oscillations can then lead to significant 
 flavor conversions if there exist  solutions with complex
 $\Omega$ or/and $\mathbf{K}$.

Furthermore, 
if the neutrino and antineutrino distributions 
 possess axial symmetry, 
 it is convenient  to  integrate
neutrino quantities over $\phi_\nu$ and 
work with  $\phi_\nu$-integrated quantities:
\begin{equation}
f_\nu(\theta_\nu)=\int\frac{E_\nu^2\mathrm{d}E_\nu \mathrm{d}\phi_\nu}{(2\pi)^3} f_\nu(\mathbf{p})
\label{eq:f}
 \end{equation}
and
\begin{equation}
G(\theta_\nu)=\int_0^{2\pi}\mathrm{d}\phi_\nu G_\mathbf{v}
\label{eq:G}
\end{equation}

To study fast modes in the SN environment, 
we then assumed a homogenous neutrino gas at each SN zone 
for which we adopted the local angular distributions from the
SN model. This could be justified by the fact that fast 
modes are expected to occur on scales much
smaller than that of the SN core. Thus,  one can safely 
ignore the global geometry of the core and focus on the
local physics of the problem. However, this approximation is 
valid as long as the neutrino instabilities
occur on scales well inside the considered SN zone.

It should be noted that a significant amount of backward traveling neutrinos
within the neutrino decoupling region 
 leads to the existence of   spatial fast instabilities even without ELN crossings
\cite{Izaguirre:2016gsx,  Capozzi:2017gqd, Abbar:2017pkh}.
However, to have growing instabilities, 
the existence of  temporal instability  with complex $\Omega$
 is necessary \cite{sturrock1958kinematics, Capozzi:2017gqd}.     
Therefore,  one can focus on the temporal modes 
having in mind that 
the spatial instabilities are only important to determine the absolute or convective 
nature of the growing instabilities \cite{Yi:2019hrp}.

\section{Occurence of ELN CROSSINGS IN MULTI-D SN MODELS}\label{sec:ELNcrossings}

\begin{figure*}[tbh!] 
\centering
\begin{center}
\includegraphics*[width=1.\textwidth, trim= 10 30 10 10,clip]{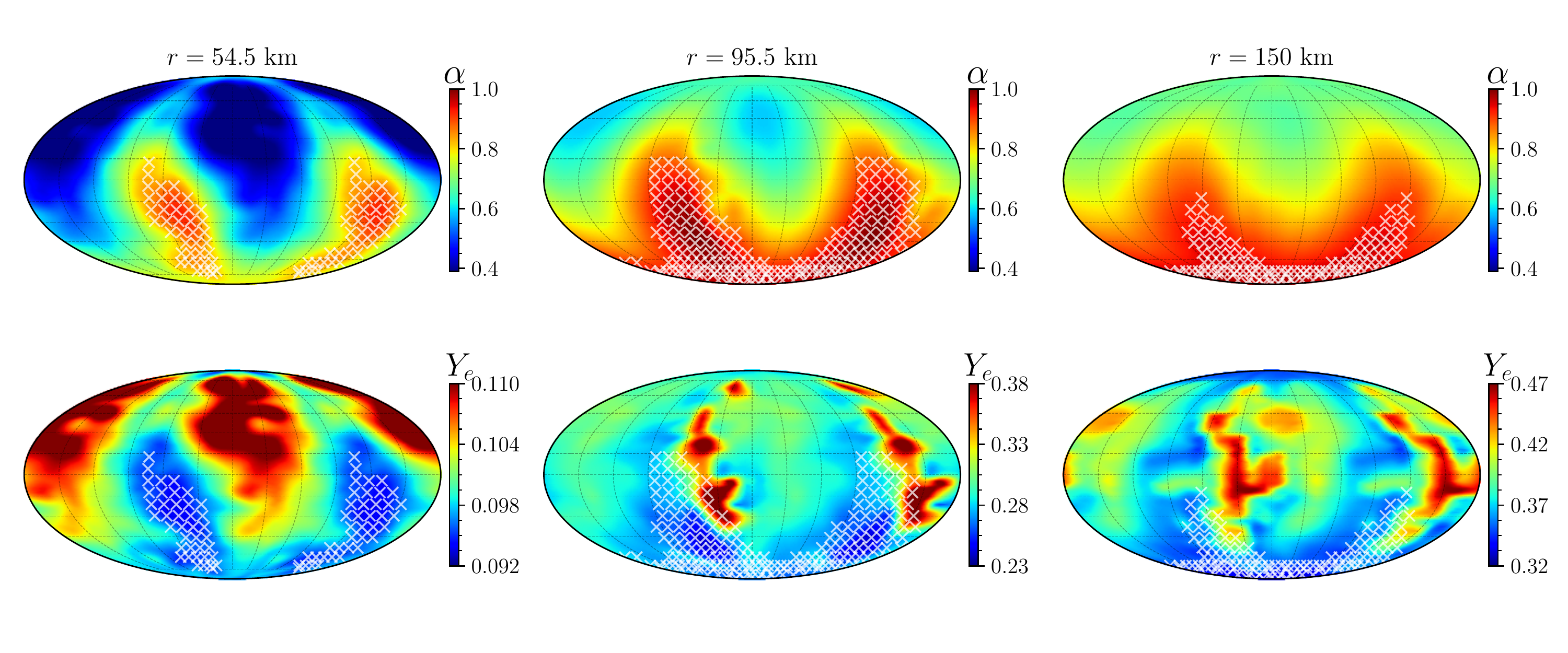}
\end{center}
\caption{
The Mollweide projection of $\alpha$
(upper panels) and the electron fraction $Y_e$ (lower panels) of the
3D  $11.2\mathrm{M}_{\odot}$
progenitor model calculations   in the  $t= 200$ ms snapshot  at $r=54.5$, 95.5 and 150 km, respectively. 
Crosses indicate the ELN crossings.
Note that the color scales may  
not be the same for different panels. At this time, 
neutrinos decouple from matter at radii $\simeq 50-70$ km,
highly depending on their flavors and energies.}
\label{fig:3D11_200}
\end{figure*}

\begin{figure*}[tbh!] 
\centering
\begin{center}
\includegraphics*[width=1.\textwidth, trim= 10 30 10 10,clip]{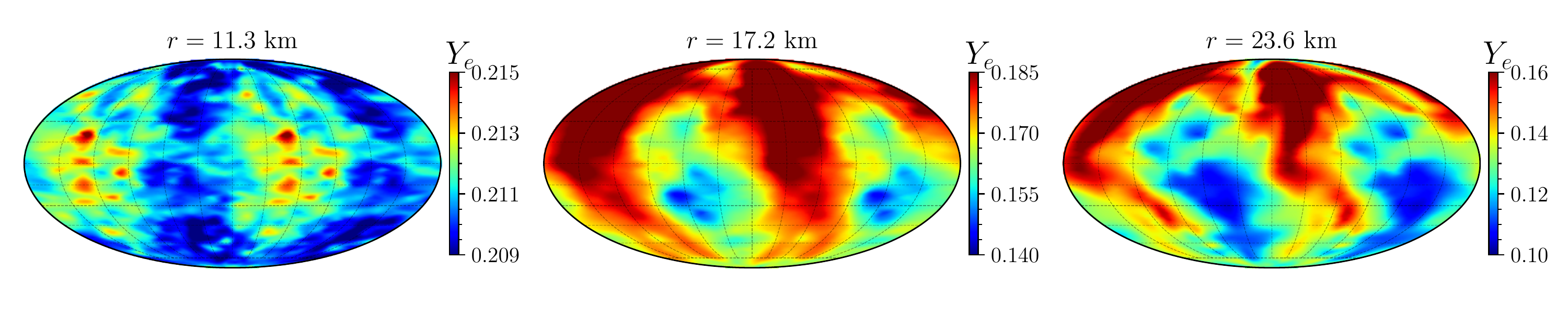}
\end{center}
\caption{
The Mollweide projection of the  electron fraction $Y_e$ of the
   3D $11.2\mathrm{M}_{\odot}$
progenitor model  calculations in the $t= 200$ ms snapshot  at $r=11.3$, 17.2 and 23.6 km, respectively. 
The pattern observed in Fig.~\ref{fig:3D11_200} first appears  very deep
inside the PNS.
Note that the color scales may  
not be the same for different panels. 
  }
\label{fig:3D11_Ye}
\end{figure*}

The neutrino distributions in this study were obtained by the calculations of  neutrino 
transport for the fixed backgrounds (density, temperature and electron fraction)
 of 2D/3D matter profiles in the supernova core.  
 The time evolution of the neutrino distributions was  followed 
 until  a stationary state for the  neutrino quantities was reached. 
 
The 
Boltzmann equation  was solved directly in the full phase space
 to obtain the neutrino energy and 
angle distributions.  
The multi-angle multi-energy neutrino transport in 2D/3D space
was carried out by using  the $S_n$ method
and the time evolution was handled by the time implicit method.  
Further details of our numerical method can be found in Refs.~\cite{Sumiyoshi:2012za, Sumiyoshi:2014qua}.  

In the neutrino transport, three neutrino species, namely, $\nu_e$, $\bar{\nu}_e$ and $\nu_x$ were handled where 
$\nu_x$ represents $\nu_\mu$ and $\nu_\tau$ and their anti-particles.  
The microphysics used in the evaluation of the neutrino transport 
is the same as the one in  Refs.~\cite{Sumiyoshi:2014qua, Abbar:2018shq} where
the neutrino reaction rates for emission, absorption, scattering and pair processes were taken 
mostly from Ref.~\cite{Bruenn:1985en} and its extensions.
  In addition, the Lattimer \& Swesty EOS
 \cite{Lattimer:1991nc}  was used to 
be consistent with the original supernova simulations.


We adopted multi-D supernova profiles
of the two progenitor models of an  $11.2\mathrm{M}_{\odot}$ (both 2D and 3D)
and a $27\mathrm{M}_{\odot}$ (only 3D) \cite{Sumiyoshi:2014qua, Takiwaki:2011db,Takiwaki:2013cqa}.
The  spatial resolutions of the Boltzmann calculations were set to be (256, 64, 1)
 and (256, 64, 32) for the numbers of spatial zones $(N_r, N_\Theta, N_\Phi)$
 in spherical coordinates
   for the 2D and 3D models, respectively, where
   a maximum radius of 2613 km 
 from the original simulations was reached.
 For the neutrino   momentum space,
 a resolution of (14, 6, 12) was employed for $(N_{E_\nu}, N_{\theta_\nu}, N_{\phi_\nu})$
 in  both 2D and 3D calculations.
 In addition, to examine the  angular convergence of  the calculations, 
 we performed a similar neutrino transport calculations
 with a higher resolution of $N_{\theta_\nu} = 36$
  for the snapshots of the 2D  $11.2\mathrm{M}_{\odot}$  progenitor model
 at $t=150$ and 200 ms.

\begin{figure*}[tbh!] 
\centering
\begin{center}
\includegraphics*[width=1.\textwidth, trim= 10 30 10 10,clip]{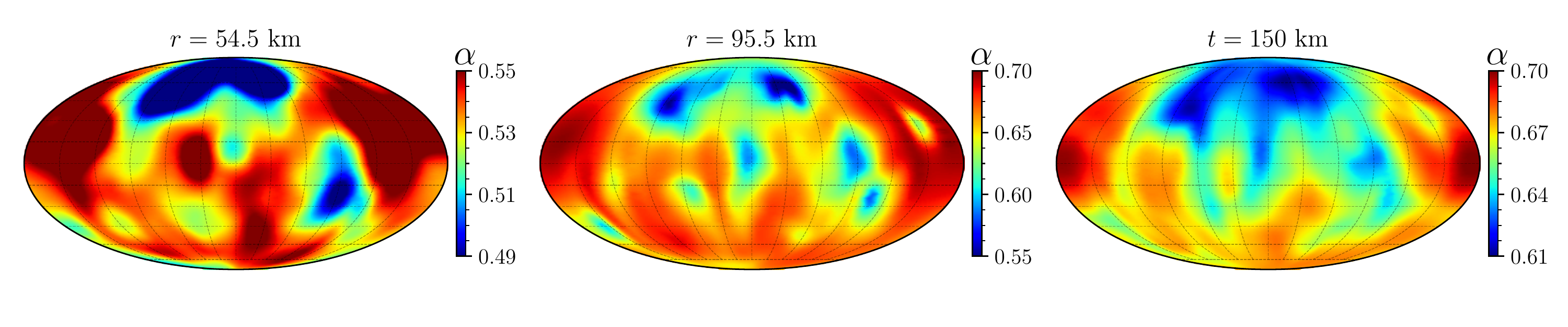}
\end{center}
\caption{
The Mollweide projection of $\alpha$ of the
3D  $27\mathrm{M}_{\odot}$
progenitor model calculations in the  $t= 200$ ms snapshot  at $r=54.5$, 95.5 and 150 km, respectively. 
Note that the color scales are  
not  the same for different panels. 
  }
\label{fig:3D27}
\end{figure*}

For the  $11.2\mathrm{M}_{\odot}$ progenitor model, three representative snapshots 
 at $t=100$, 150 and 200 ms  from the original 2D and 3D simulations 
 were selected
 for which  the Boltzmann  equations
were solved. As  mentioned in Ref.~\cite{Abbar:2018shq},
ELN crossings within/above the decoupling region were
 only observed in the snapshot at $t=200$ ms 
in  the 2D calculations
 whereas 
 they were found 
in all of the time snapshots in the 3D calculations. 
For instance, in the snapshot at $t=200$ ms, 
the ELN crossings above the neutrinosphere appear first at 
$r \simeq 46$ km within a relatively narrow region in 
the southern hemisphere (Fig.~\ref{fig:3D11_200}).
As the radius increases, the crossings zone expands initially
 because the neutrino distributions get more 
 peaked in the forward direction. However, it
 gets narrower again at larger radii and disappears at $r \sim 200$ km.
As will be discussed later, this disappearance of the ELN crossings could be an
 artificial result of the limited angular resolution of the neutrino transport calculations. 


 As pointed out in Ref.~\cite{Abbar:2018shq}, the ELN crossings zones appear to be correlated with 
the zones where the $\nu_e$-$\bar{\nu}_e$ asymmetry parameter, 
\begin{equation*}
\alpha = \frac{n_{\bar\nu_e}}{n_{\nu_e}} ,
\end{equation*}
is close to 1.
This, indeed, is not coincidental and  can be 
 understood as follows.
 In the SN environment, 
 the angular distribution of $\bar{\nu}_e$ is normally more peaked in the forward direction 
than that of $\nu_e$ because they decouple at smaller radii.
One then might be tempted to assume that 
the occurrence of ELN crossings  in  CCSNe is inevitable. 
However, 
if the $\nu_e$-$\bar{\nu}_e$ asymmetry is large,
the occurrence of ELN crossings could be remarkably  suppressed. 
In fact, a larger $\nu_e$-$\bar{\nu}_e$ asymmetry means a more
separated $\nu_e$ and $\bar{\nu}_e$ angular distributions
and therefore, less chance for the occurrence of ELN crossings.
Thus, ELN crossings are more
likely to occur within the zones with small $\nu_e$-$\bar{\nu}_e$ asymmetries
($\alpha$ close to 1).

\begin{figure*}[tbh!] 
\centering
\begin{center}
\includegraphics*[width=1.\textwidth,trim = 10 10 10 10,clip]{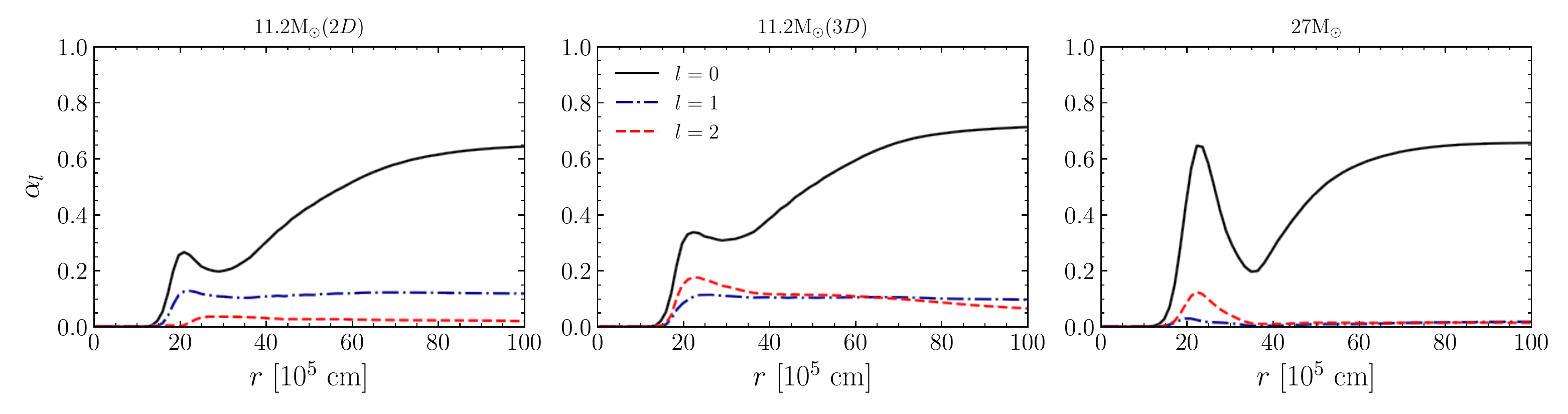}
\end{center}
\caption{
Radial evolution of the multipole components of  
$\alpha$ Eq.~\eqref{Eq:Ylm} for the $t=200$ ms snapshots.
 In  2D and 3D $11.2\mathrm{M}_{\odot}$ progenitor models, strong dipole and quadrupole components in $\alpha$ exist; while  in the  $27\mathrm{M}_{\odot}$ one, these  are only found inside the PNS.
  }
\label{fig:Ylm}
\end{figure*}

\begin{figure*}[tbh!] 
\centering
\begin{center}
\includegraphics*[width=1.\textwidth, trim= 5 10 10 10,clip]{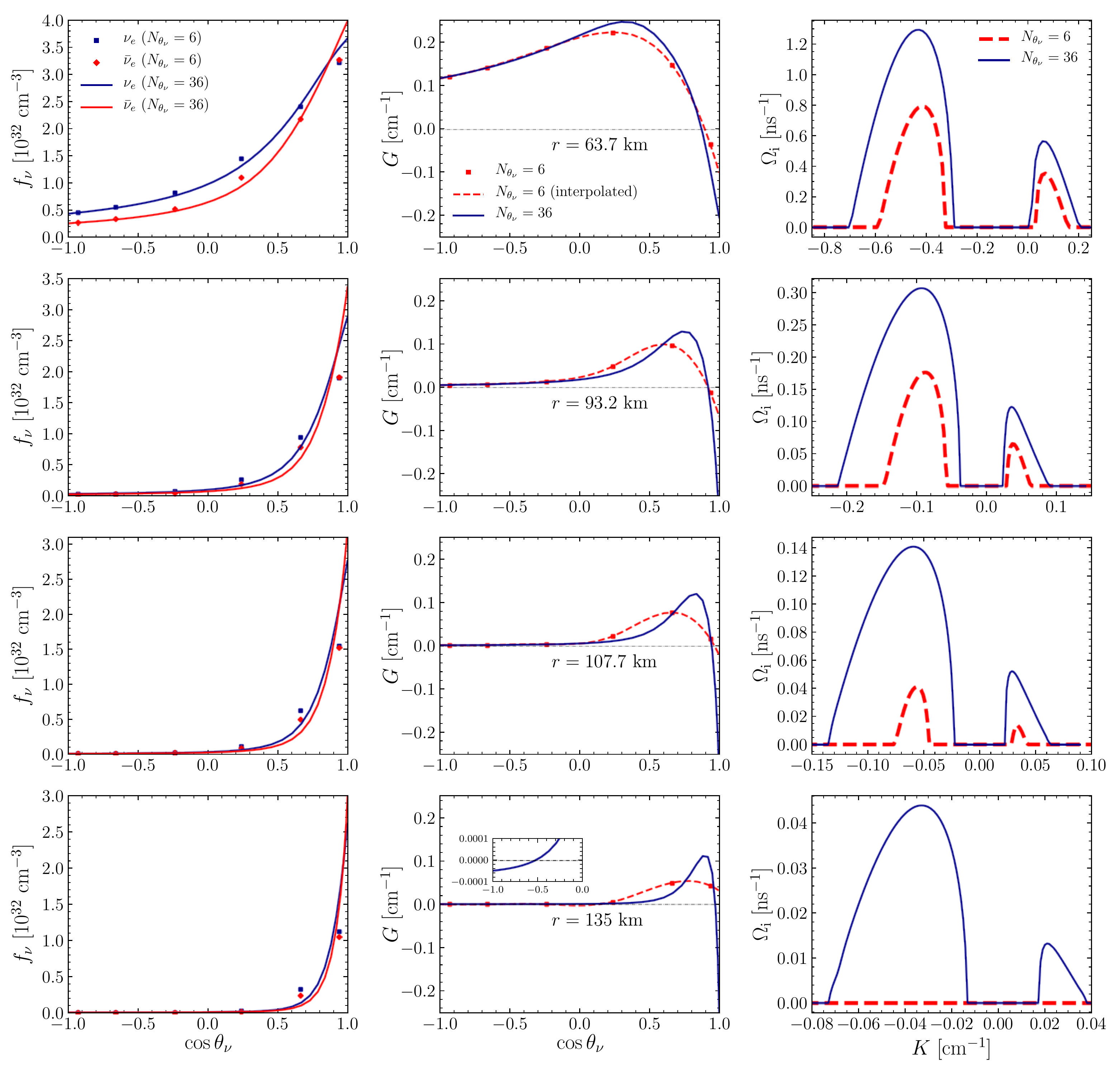}
\end{center}
\caption{Angular distributions of neutrinos $f_\nu$ (left),
ELN (middle), as  functions of the angular variable $\theta_{\nu}$, 
and the corresponding eigenvalues $\Omega_\mathrm{i}$ (right panels) Eqs.(\ref{Eq:linear}-
\ref{Eq:exp}), as functions of the real wave number $K$.  The results correspond to fast neutrino oscillation
 modes propagating in the radial direction, 
at the four different radii $r= 63.7, 93.2, 107.7$ and $135$ km. 
The angular distributions are extracted from the spatial point
with $\cos\Theta = 0.99$ in the 2D model in the $t=200$ ms snapshot.
  The zoomed-up subplot indicates the shallow crossing at  $\cos\theta_\nu \simeq -0.56$
  in the calculation with $N_{\theta_{\nu}} = 36$.
  A similar shallow crossing exists in the calculations with $N_{\theta_{\nu}} = 6$ but
  the  low angular resolution does not allow for a definite recognition of it.
  At this time, the neutrinospheres of different flavors are located
 at radius $\sim 50-70$ km (see Fig.~4 of Ref.~\cite{Abbar:2018shq}).
  }
\label{fig:G}
\end{figure*}

\begin{figure}[tbh!] 
\centering
\begin{center}
\includegraphics*[width=.4\textwidth, trim= 5 15 10 10,clip]{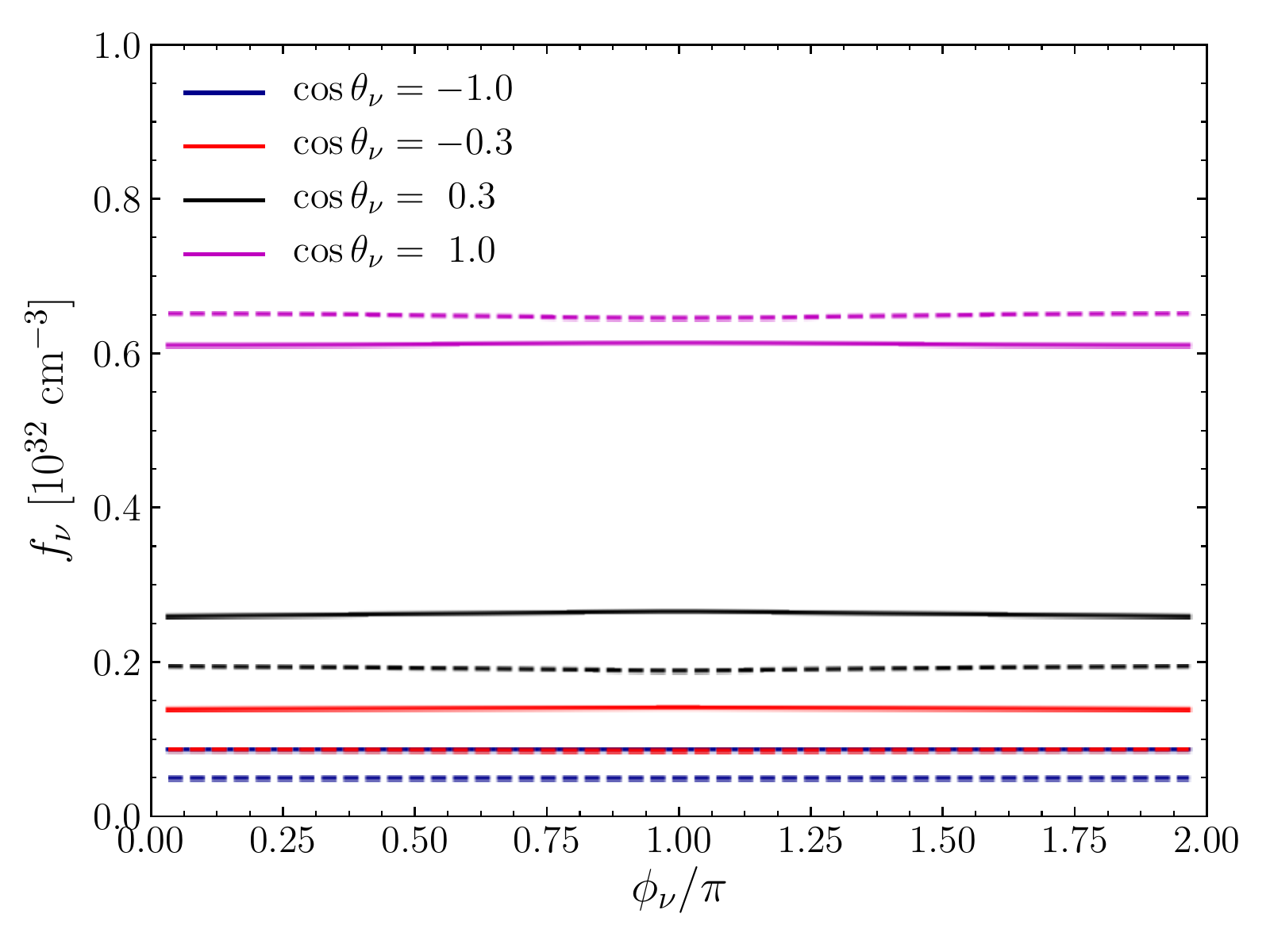}
\end{center}
\caption{
Neutrino angular distributions
as functions of $\phi_\nu$
for $\nu_e$ (solid) and $\bar\nu_e$ (dashed) and at
different $\cos\theta_\nu$ values (Eq.~(\ref{Eq:ang})). The results are 
obtained from the 2D calculations of the $t= 200$ ms $11.2\mathrm{M}_{\odot}$
progenitor model,  
for the spatial zone at $r=61.8$ km and $\cos\Theta=0.98$.
The  asymmetry in $\phi_\nu$ turns out to be  small for 
the spatial zones with ELN crossings.}
\label{fig:phi}
\end{figure}

The pattern in $\alpha$ (and the neutrino distributions) is obviously  (anti)correlated  with 
a similar pattern in the electron fraction $Y_e$ (Fig.~\ref{fig:3D11_200}).
 Fig.~\ref{fig:3D11_Ye} shows that it first appears very deep inside the PNS. 
 As pointed out in Ref.~\cite{Tamborra:2014aua}, in the case of LESA 
convectional flows in the PNS could generate asymmetries in the neutrino fluxes. 
 In our 3D models, the pattern in $Y_e$ 
 shows up first at $r \sim 11$ km and becomes more distinct at larger radii.
 The  pattern  in $Y_e$ can then be associated with  
a similar pattern in $\alpha$ because more (less) $\nu_e$'s ($\bar{\nu}_e$'s)
 are emitted  where $Y_e$ is larger. 

 For the  $27\mathrm{M}_{\odot}$ progenitor model, three representative snapshots 
 at $t=150$, 200 and 250 ms  from the original 3D simulations were selected.
 In contrast to the case of the  $11.2\mathrm{M}_{\odot}$ progenitor model,
 no ELN crossings above the neutrinosphere  were found in 
 this case. 
 A few representative 
 Mollweide projections of  $\alpha$  at different radii
 (corresponding to the ones in Fig.~\ref{fig:3D11_200} for the
  $11.2\mathrm{M}_{\odot}$ progenitor model)
  are plotted in Fig.~\ref{fig:3D27}.
 Although it shows some slight  spatial variations,
the value of $\alpha$ within/above the neutrino decoupling region
is always  much smaller than one in this model which in turn
provides a little chance for the occurrence of ELN crossings.

 During the early stages of a CCSN, (the average of) $\alpha$ tends to be 
 relatively small. This 
  can seriously  hinder
the occurrence of ELN crossings in the SN environment.
In particular, no ELN crossings within the shock region have been found 
 in 1D SN models so far \cite{Tamborra:2017ubu, Shalgar:2019kzy}
 except in the pre-shock SN region \cite{Morinaga:2019wsv}.
 However, the situation can be different  in multi-D SN models
 where the neutrino distributions can be spatially asymmetric due to multi-D hydrodynamics.
  In fact, although the average value of $\alpha$
 is thought to be similar in 1D and multi-D SN models, 
 the existence of multipole structures in $\alpha$ in the latter can
 allow for regions with large $\alpha$'s which increases
 the chance for the occurrence of ELN crossings. Fig.~\ref{fig:Ylm} presents
 the $l=0,1,2$ 
  multipole components of the spherical harmonics decomposition
 of  $\alpha$, defined as
 \begin{equation}
 \alpha_l = \bigg(   \sum_{m=-l}^l \left| \int \frac{\mathrm{d}\Omega}{\sqrt{4\pi}} \ Y^*_{lm}(\Theta,\Phi) \alpha(\Theta,\Phi) \right|^2     \bigg)^{1/2}.
 \label{Eq:Ylm}
 \end{equation} 
 As one can see, both the
  2D and the 3D $11.2\mathrm{M}_{\odot}$ progenitor models show
strong dipole and quadrupole components in $\alpha$. On the other hand
  these are weak in the $27\mathrm{M}_{\odot}$ progenitor model, 
except for a small region inside the PNS where ELN crossings exist
  (see Sec.~\ref{sec:PNS}).
 This  provides  an explanation  on why there is no ELN crossings within/above the neutrino decoupling
  region for this model.
 Furthermore, while in the 2D calculations the dipole component is dominant in all of the snapshots,
 it is the quadrupole  term that is stronger, at least at smaller radii, in all 
 of the 3D snapshots. 

 \

A few representative $\phi_\nu$-integrated angular distributions of  $\nu_e$, $\bar{\nu}_e$ 
and the corresponding ELN (Eqs.~\eqref{eq:f} and \eqref{eq:G}) 
 are presented in Fig.~\ref{fig:G}
for the  $t=200$ ms  2D $11.2\mathrm{M}_{\odot}$ progenitor model.
The angular distributions with $N_{\theta_{\nu}} = 6$ and  $N_{\theta_{\nu}} = 36$ are in 
a relatively good agreement. The only exception is
at larger radii, where
the calculations with smaller number of angle bins 
 fails (expectedly) to capture the angular
structures at small emission angles. 
In fact, 
as the radius gets larger, 
the ELN crossings get narrower 
because the neutrino angular distributions
 become more forward peaked. Therefore, higher angular resolution 
is needed to capture them.
This implies that the disappearance of the ELN crossings at larger radii
could be an artefact of the  low angular resolution of the
neutrino  transport calculations.  
The average neutrino quantities are in much better
 agreement between the two calculations with different resolutions,
  the values of $n_\nu$ and $\alpha$ differing
at most by $2-3\ \%$.

 Moreover, we noticed that 
 there are cases where the ELN distributions exhibit
 two crossings. This phenomenon  seemingly
 occurs at larger radii where the neutrino
 angular distributions are highly peaked 
 in the forward direction. As  shown in 
 Fig.~\ref{fig:G}, at r = 135 km there is a first shallow crossing at negative values of $\cos\theta_\nu$ and a second one at  $\cos\theta_\nu \simeq 1$.

Besides the analysis of the neutrino angular distributions, 
 we  performed a linear stability analysis and solved Eq.~\eqref{Eq:linear}
to find the growth rates of the unstable modes
(Eq.~\eqref{Eq:exp})
at each spatial zone. 
 The calculations assume axial symmetry for 
the neutrino gas. 
 In fact, 
the neutrino angular distributions 
\begin{equation}\label{Eq:ang}
f_\nu(\theta_\nu,\phi_\nu)=\int\frac{E_\nu^2\mathrm{d}E_\nu} {(2\pi)^3} f_\nu(\mathbf{p}),
\end{equation}
are highly symmetric in $\phi_\nu$ 
 at all of the spatial zones for  which  ELN crossings were observed, as illustrated by 
Fig.~\ref{fig:phi}. 
Overall, the deviation from axial symmetry
in the zones with ELN crossings
 turns out to be smaller than a few percent
in both 2D and 3D models.
However,  one should keep in mind that this
can be changed for rotating models where the angular
distributions can be quite asymmetric in $\phi_\nu$ \cite{Harada:2018ubo}.
As mentioned earlier,
the temporal instability
is necessary to have 
 growing perturbations. 
 The complex eigenvalues associated with the temporal 
fast modes are plotted in Fig.~\ref{fig:G}.
The corresponding exponential growth rates
can be as large as several e-folds per nanoseconds and 
tend to be larger for the wider ELN crossings.

Our results suggest that 
even the more accessible calculations with 
low angular resolution might provide
 a quite reliable estimate of the presence of ELN crossings and correspondingly fast modes.
 This is corroborated by the fact that 
 we did not find false ELN crossings 
 in the calculations with lower angular resolution and the associated instability growth
 rates provided a reasonable estimate of the results
based on higher angular resolution. Obviously, one cannot exclude that 
in a dynamic self-consistent CCSN simulation,
 the prompt convection might be more suppressed
 if the angular resolution is too low \cite{Nagakura:2017mnp}.

\begin{figure*}[tbh!] 
\centering
\begin{center}
\includegraphics*[width=1.\textwidth, trim= 10 30 10 10,clip]{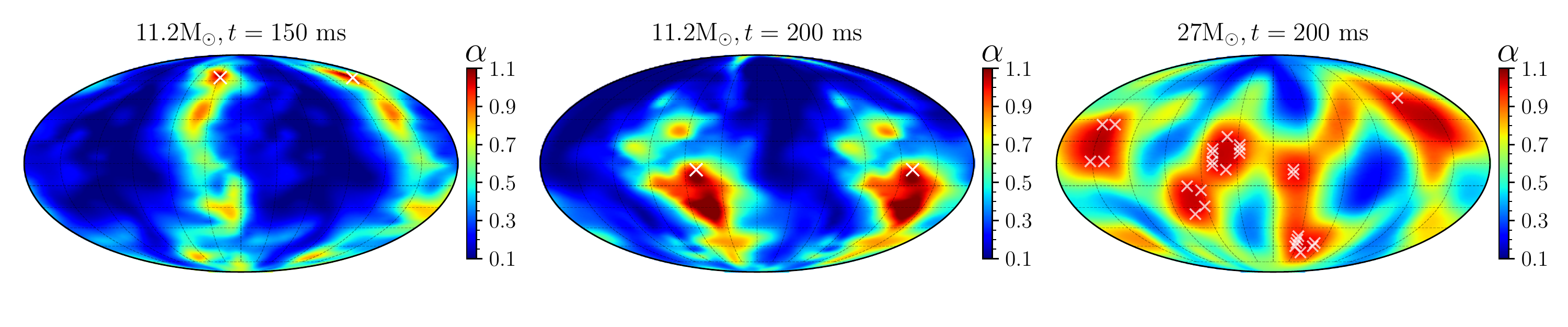}
\end{center}
\caption{
The Mollweide projection of $\alpha$
of the 3D calculations  at $r=23.6$ km  in different time snapshots.
The few crosses indicate ELN crossings inside the PNS. Note
that the crossings inside the PNS are not as abundant as
the ones within the decoupling region and they only exist within the zones where $\alpha$ is extremely close to one.
  }
\label{fig:PNS}
\end{figure*}

\begin{figure*}[tbh!] 
\centering
\begin{center}
\includegraphics*[width=1.\textwidth, trim= 10 15 10 10,clip]{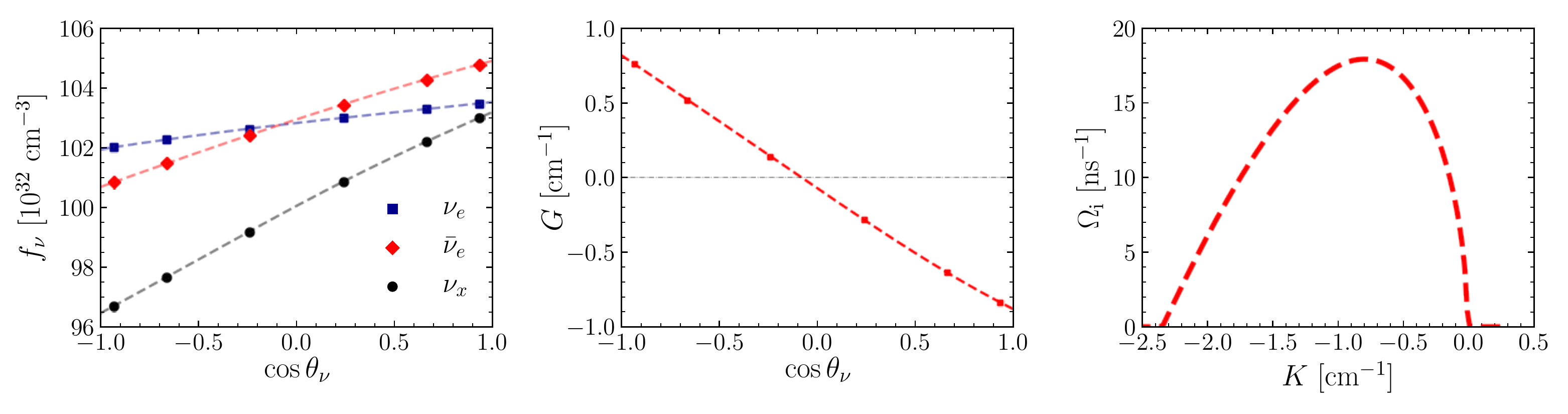}
\end{center}
\caption{ Angular distributions, as functions of $\theta_{\nu}$,
for $\nu_e$, $\bar\nu_e$ and $\nu_x$
(left), ELN (middle) and the 
corresponding eigenvalues $\Omega_\mathrm{i}$  (right panel), 
as a function of the real wave number $K$.  Interpolated values are also shown (dashed lines).
 The results are for the fast
 modes propagating in the radial direction, 
 at the spatial zone with
$r=26.4$, $ \cos\Theta = 0.40$ and $\Phi = 0.72\pi$
deep inside the PNS, for which an ELN crossing exists.
The calculations are for 3D $11.2\mathrm{M}_{\odot}$
progenitor model and the $t= 200$ ms snapshot.
Note that the neutrino angular distributions are very close to each
other and highly  non-degenerate, with $n_{\bar\nu_e}/n_{\nu_e} = 1.001$
and $ n_{\nu_x}/ n_{\nu_e} = 0.972$. }
\label{fig:PNS-fnu}
\end{figure*}

\begin{figure*}[tbh!] 
\centering
\begin{center}
\includegraphics*[width=1.\textwidth, trim= 10 30 10 10,clip]{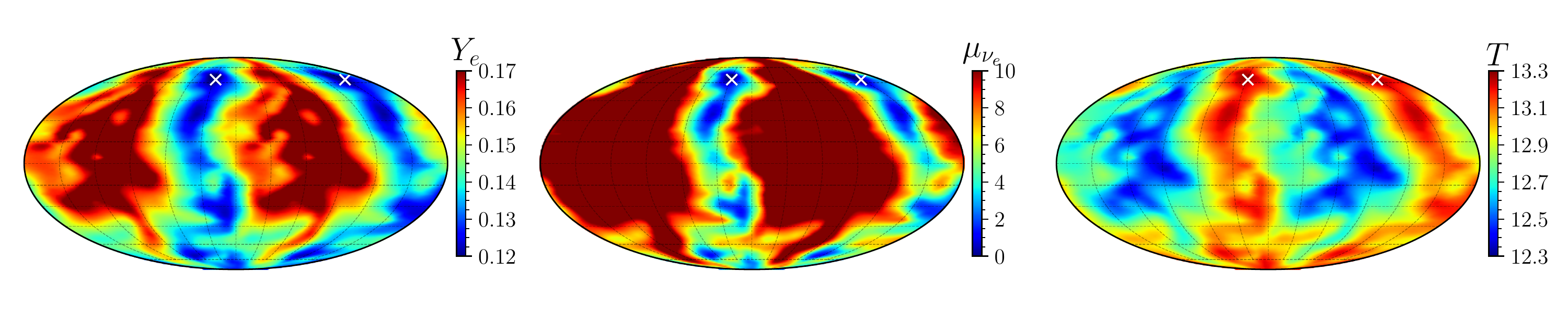}
\end{center}
\caption{
The Mollweide projection of the electron fraction $Y_e$,
the electron neutrino chemical potential $\mu_{\nu_e}$
and the temperature $T$
of the 3D $11.2\mathrm{M}_{\odot}$ progenitor model calculations  at $r=23.6$ km 
in the $t=150$ ms snapshot corresponding to the very left panel
in Fig.~\ref{fig:PNS}.
$\mu_{\nu_e}$
and $T$ are both in MeV.
 }
\label{fig:mu_nu}
\end{figure*}

\begin{figure}[tbh!] 
\centering
\begin{center}
\includegraphics*[width=.48\textwidth, trim= 5 0 10 5,clip]{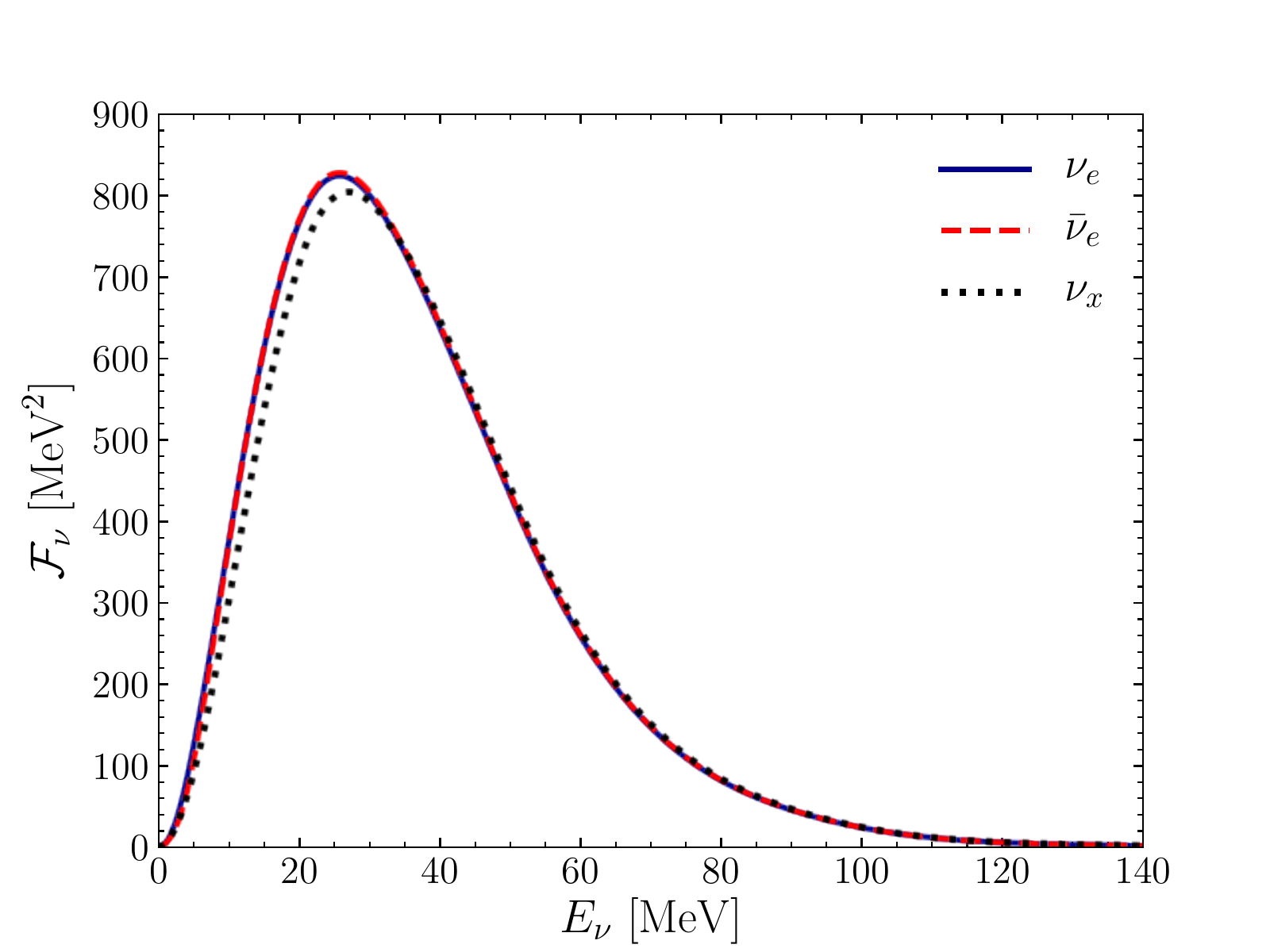}
\end{center}
\caption{
The neutrino energy distributions 
 at the spatial point corresponding to the one in Fig.~\ref{fig:PNS-fnu}.}
\label{fig:eng}
\end{figure}

\section{ELN crossings  inside the PNS } \label{sec:PNS}
\noindent
 Apart from the ELN crossings  occurring 
 within/above the decoupling region of neutrinos, there  is another 
 class of ELN crossings that can appear in deep regions inside
 the PNS. In our calculations, such ELN crossings 
exist  at radii in the range $20-28$ km.

  For the 2D $11.2\mathrm{M}_{\odot}$ progenitor model, only two ELN crossings at $\sim 25$ km
 exist in the snapshot at $t=100$ ms, while
 they appear in all 3D snapshots.
 As for the  $27\mathrm{M}_{\odot}$  progenitor
 model, the ELN deep crossings  are present
  in the $t=150$ and 200 ms snapshots  (Fig.~\ref{fig:PNS}).
Note that there exist no ELN crossings within the decoupling region
 for this progenitor model. 

 Since the neutrino angular distributions are
 significantly  isotropic inside the PNS, the deep crossings can
 only  exist if $\alpha$ is extremely  close to one, i.e.,
 the $\nu_e$ and $\bar{\nu}_e$ angular distributions are very close to each other 
 (Fig.~\ref{fig:PNS-fnu}). Note that if $\alpha$ has values both below and
 above 1, it is very likely to have this sort of ELN crossings
 provided that the  spatial  resolution is high enough. 
 With this in mind, it is safe to say that the key question here is whether such hot spots (as in Fig.~\ref{fig:PNS}) do actually
 show up in  more realistic SN simulations.
 It should be also noted that such ELN crossings do not necessarily occur
 at $\cos\theta_\nu \simeq 1$.
 
 The reason for having  SN zones for which $\alpha \simeq 1$ can be
 understood as follows. As shown in
 Fig.~\ref{fig:mu_nu},
 the neutrino gas is very non-degenerate ($\mu_{\nu}/T \simeq 0$) at the ELN crossing points
 inside the PNS. 
   In fact, the chemical potentials of
$\nu_e$ and $\bar{\nu}_e$ become similar at these points, i.e., $\mu_{\nu_e}
\simeq \mu_{\bar{\nu}_e} \simeq 0$ ($\mu_{\nu_e} = \mu_{\mathrm{p}} - \mu_{\mathrm{n}} + \mu_{e}$).
This happens because  $\mu_{\mathrm{p}} - \mu_{\mathrm{n}} $ 
can  almost cancel $\mu_{e}$ at these SN zones where $Y_e$ is minimum\footnote{
We would like to thank Shoichi Yamada for valuable conversation about this
possiblity.}.
 Thus, a  correlation exists between $Y_e$ and $\mu_{\nu_e}$.
In addition, the temperature has its maximum value at these points.
As mentioned in Ref.~\cite{Janka:2016fox}, 
this (anti)correlation between $Y_e$ and $T$ can be explained by noting that
in spite of the asymmetric neutrino distributions, 
the density and pressure tend to maintain their spherical symmetry. 
This simply arises from the sphericality of the gravitational potential
which mainly governs the variations of these quantities. Therefore,
one should expect higher temperatures within the zones where 
$Y_e$ is lower to keep the combination of  thermal  and
 electron degeneracy pressures constant ($p = p(\rho,T,Y_e)$).

 Although the existence  of the  deep ELN crossings 
 surprisingly allows for the occurrence  of fast modes
 inside the PNS,
there are two essential points that should be kept in mind.
 Firstly,  one 
 can indeed observe  large flavor conversion rates
  for all the SN  zones where   
 $\alpha$ is extremely close to one,  even for those the
 ELN crossings are absent. 
 For example,  in an isotropic homogeneous 
 monoenergetic neutrino gas initially consisting  of  $\nu_e$ and $\bar{\nu}_e$ (bipolar model),
 for $|\alpha-1| \ll 1$, the exponential growth rate  governing flavor evolution
 is \cite{Duan:2007fw, mangano:2014zda}
  \begin{equation}
\Omega_\mathrm{i} \approx  \sqrt{2\mu\omega},
 \end{equation}
for
 \begin{equation}
 \frac{2}{(1+\sqrt{\alpha})^2} < \mu/\omega < \frac{2}{(1-\sqrt{\alpha})^2},
 \end{equation}
 for
 $\eta=1$\footnote{In the case of slow modes, the  exponential growth rate is 
  $\sim  \sqrt {\omega \mu}$. However, unless $\alpha$ is extremely close to one, 
  $\mu \sim \omega$ 
  which means $\Omega_\mathrm{i} \sim \omega$. One can
  observe a similar instability for $\eta=-1$ by breaking the axial/spatial symmetry.}.
 This implies that  even in the bipolar model (in the absence of ELN crossings), 
 there can exist unstable 
 slow modes  on scales  $\propto G_{\mathrm{F}}^{-1/2} n_\nu ^{-1/2}$ 
   at spatial zones where $\alpha \simeq 1$.
  Such scales  can be $\lesssim 1$ m 
  (compare with fast modes occuring  on scales $\sim G_{\mathrm{F}}^{-1} n_\nu ^{-1} \lesssim  10$ cm)
  for these SN zones inside the PNS.
This is     much shorter than the collisional scales ($\lesssim  \mathcal O  (10^2)$ m) therein. 
Therefore, having large neutrino flavor conversion rates
for non-degenerate SN zones is not unique to fast modes.
   
 Secondly, the neutrino distributions are significantly similar for all neutrino species within these zones 
meaning that not only $n_{\nu_e} \simeq n_{\bar{\nu}_e} \simeq n_{\nu_x}$ (Fig.~\ref{fig:PNS-fnu}),
but also the energy distributions
 \begin{equation}
\mathcal{F}_\nu(E_\nu)=\int \mathrm{d}\Gamma_{\mathbf{v}} E_\nu^2  f_\nu(\mathbf{p}),
 \end{equation}
 are very similar for
different neutrino species  (Fig.~\ref{fig:eng}).
Consequently,   neutrino oscillations should not have significant impact
on the flavor content of the neutrino gas.

\section{Discussion and conclusion}

To assess the possibility of the occurrence of ELN crossings and the
associated fast neutrino flavor conversion
modes in the SN environment,  
we have studied neutrino angular distributions obtained by
solving the Boltzmann equations for  fixed  matter profiles of some
representative  snapshots   during the post-bounce phase of CCSNe
in 2D and 3D  for  an $11.2\mathrm{M}_{\odot}$ and a $27\mathrm{M}_{\odot}$ (only 3D)  
progenitor  models. 

For the $11.2\mathrm{M}_{\odot}$  progenitor model, 
ELN crossings were observed in both 2D and 3D models.
It turns out that 
they tend to occur within the SN zones where the 
 $\nu_e-\bar{\nu}_e$ asymmetry parameter, $\alpha$,
 is very close to one,  as in Ref.\cite{Abbar:2018shq}.
 Unlike the 1D SN modes in which $\alpha$ is 
 thought to be globally small during the early stages  
 of a CCSN, 
  zones with large $\alpha$'s could exist in multi-D SN models in the presence of 
 a spatially asymmetric neutrino distributions due to multi-D hydrodynamics.
 The reason is that in spite of having similar average values of
 $\alpha$ to 1D models, multi-D SN models can feature
 larger $\alpha$'s in  zones where $\alpha$ is significantly larger  than its averaged value. 

 The pattern in $\alpha$ is  (anti)correlated with a similar pattern
 in $Y_e$ (and $T$) that appears very deep inside the PNS and is thought to
 be caused by the convectional flows therein. 
 This is closely related to LESA  
  \cite{Tamborra:2014aua, Vartanyan:2019ssu, Sugiura:2019xuv, Walk:2019ier, Glas:2018vcs,
 OConnor:2018tuw, Janka:2016fox, Tamborra:2014hga, Nagakura:2019evv}
 in which strong dipole  structure exists in the neutrino distributions
 which can be preceded by strong higher multipole structures at earlier times.
 Our study suggests that if the multipole structures (either  dipole at
 later times or higher multipoles at earlier times) are strong enough,
 they can lead to the existence of zones with significantly large $\alpha$'s
 and a good chance for the occurrence of ELN crossings.
  However,     at this point in time it is not very clear how the fast
 modes propagate from these zones to other regions. 
 Note also that such asymmetric structures  
 in $\alpha$ do not necessarily need to come from LESA.
 In general any significant asymmetry in the neutrino distributions,
 self-sustained with relatively constant direction (as in LESA) or not,  can
 increase the chance for the occurrence of ELN crossings in the SN
 environment.

 To see how increasing the neutrino angular resolution 
can affect the occurrence of ELN crossings, we 
performed  calculations with higher 
angular resolution with $N_{\theta_\nu}=36$ in 2D
for our $11.2\mathrm{M}_{\odot}$  progenitor model.
It turns out that not only the calculations with 
lower resolution do not produce any false 
ELN crossings, but also they can provide a relatively
good estimate of the flavor conversion  rates in a more 
reliable calculation with higher angular resolution.
This is intriguing since the calculations with higher 
resolution are barely accessible in multi-D SN models.

 For the $27\mathrm{M}_{\odot}$  progenitor model, 
 no ELN crossings within the neutrino decoupling region were found.
 Although there exists a  quadrupole structure in $\alpha$,
 it is still too weak to result in the existence of regions
 with  large $\alpha$'s. This comes as no surprise
 since the evolution of the multipole structures in neutrino
 emission   depends sensitively on a number factors such as
 the mass of the progenitor,  
 the treatment of  neutrino transport and even the employed EOS
 (see, e.g., Fig.~5 in Ref.~\cite{Janka:2016fox}).
 
  We would like to point out that, no
 ELN crossings were found 
  in a  recent self-consistent 2D SN simulations
 using the Furusawa’s EOS \cite{Furusawa:2011wh}, as opposed to the Lattimer \& Swesty EOS  
 used in our calculations.  This result is not inconsistent with
 the present study nor with Ref.~\cite{Abbar:2018shq}, since
 convective  motion, crucial to the existence
 and evolution of neutrino distribution asymmetries, 
 are  much weaker in the simulations with Furusawa’s EOS \cite{Nagakura:2017mnp}. 
 Therefore, one may expect to observe weaker  distribution
 asymmetries in the simulation with the latter EOS.

Apart from the ELN crossings appearing within/above the neutrino decoupling region,
 our results show that there exist a class of ELN crossings which can occur
in  deep regions, well inside the PNS.
We observed this sort of ELN crossings in both of 
the  progenitor models we investigated.
 Because the neutrino angular distributions
are extremely isotropic below the neutrinosphere, these ELN crossings can exist only
within the zones where 
$\alpha \simeq 1$.

Needless to say, if ELN crossings and their corresponding  fast modes
 actually occur in the SN environment and lead to significant flavor conversion, they can have 
profound consequences
for the SN physics.


\section*{Acknowledgments}
S.A. and M.C.V.  acknowledge support from ”Physique 
fondamentale et ondes gravitationelles” (PhysFOG) of the Observatoire de Paris.
H.D. is supported by US DOE NP grant No. DE-SC0017803 at UNM.
This work is supported by
Grant-in-Aid for Scientific Research
(26104006, 15K05093, 19K03837)
and 
Grant-in-Aid for Scientific Research on Innovative areas 
"Gravitational wave physics and astronomy:Genesis"
(17H06357, 17H06365)
from the Ministry of Education, Culture, Sports, Science and Technology (MEXT), Japan 
For providing high performance computing resources, 
Computing Research Center, KEK, 
JLDG on SINET4 of NII, 
Research Center for Nuclear Physics, Osaka University, 
Yukawa Institute of Theoretical Physics, Kyoto University, 
Nagoya University, 
Hokkaido University, 
and 
Information Technology Center, University of Tokyo are acknowledged. 
This work was partly supported by 
research programs at K-computer of the RIKEN AICS, 
HPCI Strategic Program of Japanese MEXT, 
"Priority Issue on Post-K computer" (Elucidation of the Fundamental Laws and Evolution of the Universe)
and 
Joint Institute for Computational Fundamental Sciences (JICFus).


\bibliographystyle{elsarticle-num}
\bibliography{fast_modes}

\begin{thebibliography}{10}
\expandafter\ifx\csname url\endcsname\relax
  \def\url#1{\texttt{#1}}\fi
\expandafter\ifx\csname urlprefix\endcsname\relax\def\urlprefix{URL }\fi
\expandafter\ifx\csname href\endcsname\relax
  \def\href#1#2{#2} \def\path#1{#1}\fi

\bibitem{Colgate:1966ax}
S.~A. Colgate, R.~H. White, {The Hydrodynamic Behavior of Supernovae
  Explosions}, Astrophys. J. 143 (1966) 626.
\newblock \href {http://dx.doi.org/10.1086/148549} {\path{doi:10.1086/148549}}.

\bibitem{Qian:1996xt}
Y.~Z. Qian, S.~E. Woosley, {Nucleosynthesis in neutrino driven winds: 1. The
  Physical conditions}, Astrophys. J. 471 (1996) 331--351.
\newblock \href {http://arxiv.org/abs/astro-ph/9611094}
  {\path{arXiv:astro-ph/9611094}}, \href {http://dx.doi.org/10.1086/177973}
  {\path{doi:10.1086/177973}}.

\bibitem{Bethe:1984ux}
H.~A. Bethe, J.~R. Wilson, {Revival of a stalled supernova shock by neutrino
  heating}, Astrophys. J. 295 (1985) 14--23.
\newblock \href {http://dx.doi.org/10.1086/163343} {\path{doi:10.1086/163343}}.

\bibitem{Janka:2012wk}
H.-T. Janka, {Explosion Mechanisms of Core-Collapse Supernovae}, Ann. Rev.
  Nucl. Part. Sci. 62 (2012) 407--451.
\newblock \href {http://arxiv.org/abs/1206.2503} {\path{arXiv:1206.2503}},
  \href {http://dx.doi.org/10.1146/annurev-nucl-102711-094901}
  {\path{doi:10.1146/annurev-nucl-102711-094901}}.

\bibitem{Burrows:2012ew}
A.~Burrows, {Colloquium: Perspectives on core-collapse supernova theory}, Rev.
  Mod. Phys. 85 (2013) 245.
\newblock \href {http://arxiv.org/abs/1210.4921} {\path{arXiv:1210.4921}},
  \href {http://dx.doi.org/10.1103/RevModPhys.85.245}
  {\path{doi:10.1103/RevModPhys.85.245}}.

\bibitem{Gava:2009pj}
J.~Gava, J.~Kneller, C.~Volpe, G.~C. McLaughlin, {A Dynamical collective
  calculation of supernova neutrino signals}, Phys. Rev. Lett. 103 (2009)
  071101.
\newblock \href {http://arxiv.org/abs/0902.0317} {\path{arXiv:0902.0317}},
  \href {http://dx.doi.org/10.1103/PhysRevLett.103.071101}
  {\path{doi:10.1103/PhysRevLett.103.071101}}.

\bibitem{Horiuchi:2008jz}
S.~Horiuchi, J.~F. Beacom, E.~Dwek, {The Diffuse Supernova Neutrino Background
  is detectable in Super-Kamiokande}, Phys. Rev. D79 (2009) 083013.
\newblock \href {http://arxiv.org/abs/0812.3157} {\path{arXiv:0812.3157}},
  \href {http://dx.doi.org/10.1103/PhysRevD.79.083013}
  {\path{doi:10.1103/PhysRevD.79.083013}}.

\bibitem{Beacom:2010kk}
J.~F. Beacom, {The Diffuse Supernova Neutrino Background}, Ann. Rev. Nucl.
  Part. Sci. 60 (2010) 439--462.
\newblock \href {http://arxiv.org/abs/1004.3311} {\path{arXiv:1004.3311}},
  \href {http://dx.doi.org/10.1146/annurev.nucl.010909.083331}
  {\path{doi:10.1146/annurev.nucl.010909.083331}}.

\bibitem{Mirizzi:2015eza}
A.~Mirizzi, I.~Tamborra, H.-T. Janka, N.~Saviano, K.~Scholberg, R.~Bollig,
  L.~Hudepohl, S.~Chakraborty, {Supernova Neutrinos: Production, Oscillations
  and Detection}, Riv. Nuovo Cim. 39~(1-2) (2016) 1--112.
\newblock \href {http://arxiv.org/abs/1508.00785} {\path{arXiv:1508.00785}},
  \href {http://dx.doi.org/10.1393/ncr/i2016-10120-8}
  {\path{doi:10.1393/ncr/i2016-10120-8}}.

\bibitem{Horiuchi:2017qja}
S.~Horiuchi, K.~Sumiyoshi, K.~Nakamura, T.~Fischer, A.~Summa, T.~Takiwaki,
  H.-T. Janka, K.~Kotake, {Diffuse Supernova Neutrino Background from extensive
  core-collapse simulations of $8-100 M_\odot$ progenitors}, Mon. Not. Roy.
  Astron. Soc. 475 (2018) 1363.
\newblock \href {http://arxiv.org/abs/1709.06567} {\path{arXiv:1709.06567}},
  \href {http://dx.doi.org/10.1093/mnras/stx3271}
  {\path{doi:10.1093/mnras/stx3271}}.

\bibitem{Suwa:2019svl}
Y.~Suwa, K.~Sumiyoshi, K.~Nakazato, Y.~Takahira, Y.~Koshio, M.~Mori, R.~A.
  Wendell, {Observing Supernova Neutrino Light Curves with Super-Kamiokande:
  Expected Event Number over 10 s}, Astrophys. J. 881 (2019) 139.
\newblock \href {http://arxiv.org/abs/1904.09996} {\path{arXiv:1904.09996}},
  \href {http://dx.doi.org/10.3847/1538-4357/ab2e05}
  {\path{doi:10.3847/1538-4357/ab2e05}}.

\bibitem{Pastor:2002we}
S.~Pastor, G.~Raffelt, Flavor oscillations in the supernova hot bubble region:
  Nonlinear effects of neutrino background, Phys. Rev. Lett. 89 (2002) 191101.
\newblock \href {http://arxiv.org/abs/astro-ph/0207281}
  {\path{arXiv:astro-ph/0207281}}.

\bibitem{duan:2006an}
H.~Duan, G.~M. Fuller, J.~Carlson, Y.-Z. Qian, {Simulation of Coherent
  Non-Linear Neutrino Flavor Transformation in the Supernova Environment. 1.
  Correlated Neutrino Trajectories}, Phys. Rev. D74 (2006) 105014.
\newblock \href {http://arxiv.org/abs/astro-ph/0606616}
  {\path{arXiv:astro-ph/0606616}}, \href
  {http://dx.doi.org/10.1103/PhysRevD.74.105014}
  {\path{doi:10.1103/PhysRevD.74.105014}}.

\bibitem{duan:2006jv}
H.~Duan, G.~M. Fuller, J.~Carlson, Y.-Z. Qian, {Coherent Development of
  Neutrino Flavor in the Supernova Environment}, Phys. Rev. Lett. 97 (2006)
  241101.
\newblock \href {http://arxiv.org/abs/astro-ph/0608050}
  {\path{arXiv:astro-ph/0608050}}, \href
  {http://dx.doi.org/10.1103/PhysRevLett.97.241101}
  {\path{doi:10.1103/PhysRevLett.97.241101}}.

\bibitem{duan:2010bg}
H.~Duan, G.~M. Fuller, Y.-Z. Qian, {Collective Neutrino Oscillations}, Ann.
  Rev. Nucl. Part. Sci. 60 (2010) 569--594.
\newblock \href {http://arxiv.org/abs/1001.2799} {\path{arXiv:1001.2799}},
  \href {http://dx.doi.org/10.1146/annurev.nucl.012809.104524}
  {\path{doi:10.1146/annurev.nucl.012809.104524}}.

\bibitem{Chakraborty:2016yeg}
S.~Chakraborty, R.~Hansen, I.~Izaguirre, G.~Raffelt, {Collective neutrino
  flavor conversion: Recent developments}, Nucl. Phys. B908 (2016) 366--381.
\newblock \href {http://arxiv.org/abs/1602.02766} {\path{arXiv:1602.02766}},
  \href {http://dx.doi.org/10.1016/j.nuclphysb.2016.02.012}
  {\path{doi:10.1016/j.nuclphysb.2016.02.012}}.

\bibitem{duan:2007sh}
H.~Duan, G.~M. Fuller, J.~Carlson, Y.-Z. Qian, {Flavor Evolution of the
  Neutronization Neutrino Burst from an O-Ne-Mg Core-Collapse Supernova}, Phys.
  Rev. Lett. 100 (2008) 021101.
\newblock \href {http://arxiv.org/abs/0710.1271} {\path{arXiv:0710.1271}},
  \href {http://dx.doi.org/10.1103/PhysRevLett.100.021101}
  {\path{doi:10.1103/PhysRevLett.100.021101}}.

\bibitem{dasgupta:2009mg}
B.~Dasgupta, A.~Dighe, G.~G. Raffelt, A.~Y. Smirnov, {Multiple Spectral Splits
  of Supernova Neutrinos}, Phys. Rev. Lett. 103 (2009) 051105.
\newblock \href {http://arxiv.org/abs/0904.3542} {\path{arXiv:0904.3542}},
  \href {http://dx.doi.org/10.1103/PhysRevLett.103.051105}
  {\path{doi:10.1103/PhysRevLett.103.051105}}.

\bibitem{Galais:2011gh}
S.~Galais, C.~Volpe, {The neutrino spectral split in core-collapse supernovae:
  a magnetic resonance phenomenon}, Phys. Rev. D84 (2011) 085005.
\newblock \href {http://arxiv.org/abs/1103.5302} {\path{arXiv:1103.5302}},
  \href {http://dx.doi.org/10.1103/PhysRevD.84.085005}
  {\path{doi:10.1103/PhysRevD.84.085005}}.

\bibitem{Duan:2007bt}
H.~Duan, G.~M. Fuller, J.~Carlson, Y.-Z. Qian, {Neutrino Mass Hierarchy and
  Stepwise Spectral Swapping of Supernova Neutrino Flavors}, Phys. Rev. Lett.
  99 (2007) 241802.
\newblock \href {http://arxiv.org/abs/0707.0290} {\path{arXiv:0707.0290}},
  \href {http://dx.doi.org/10.1103/PhysRevLett.99.241802}
  {\path{doi:10.1103/PhysRevLett.99.241802}}.

\bibitem{Sawyer:2005jk}
R.~F. Sawyer, {Speed-up of neutrino transformations in a supernova
  environment}, Phys. Rev. D72 (2005) 045003.
\newblock \href {http://arxiv.org/abs/hep-ph/0503013}
  {\path{arXiv:hep-ph/0503013}}, \href
  {http://dx.doi.org/10.1103/PhysRevD.72.045003}
  {\path{doi:10.1103/PhysRevD.72.045003}}.

\bibitem{Sawyer:2015dsa}
R.~F. Sawyer, {Neutrino cloud instabilities just above the neutrino sphere of a
  supernova}, Phys. Rev. Lett. 116~(8) (2016) 081101.
\newblock \href {http://arxiv.org/abs/1509.03323} {\path{arXiv:1509.03323}},
  \href {http://dx.doi.org/10.1103/PhysRevLett.116.081101}
  {\path{doi:10.1103/PhysRevLett.116.081101}}.

\bibitem{Chakraborty:2016lct}
S.~Chakraborty, R.~S. Hansen, I.~Izaguirre, G.~Raffelt, {Self-induced neutrino
  flavor conversion without flavor mixing}, JCAP 1603~(03) (2016) 042.
\newblock \href {http://arxiv.org/abs/1602.00698} {\path{arXiv:1602.00698}},
  \href {http://dx.doi.org/10.1088/1475-7516/2016/03/042}
  {\path{doi:10.1088/1475-7516/2016/03/042}}.

\bibitem{Izaguirre:2016gsx}
I.~Izaguirre, G.~Raffelt, I.~Tamborra, {Fast Pairwise Conversion of Supernova
  Neutrinos: A Dispersion-Relation Approach}, Phys. Rev. Lett. 118~(2) (2017)
  021101.
\newblock \href {http://arxiv.org/abs/1610.01612} {\path{arXiv:1610.01612}},
  \href {http://dx.doi.org/10.1103/PhysRevLett.118.021101}
  {\path{doi:10.1103/PhysRevLett.118.021101}}.

\bibitem{Wu:2017qpc}
M.-R. Wu, I.~Tamborra, {Fast neutrino conversions: Ubiquitous in compact binary
  merger remnants}, Phys. Rev. D95~(10) (2017) 103007.
\newblock \href {http://arxiv.org/abs/1701.06580} {\path{arXiv:1701.06580}},
  \href {http://dx.doi.org/10.1103/PhysRevD.95.103007}
  {\path{doi:10.1103/PhysRevD.95.103007}}.

\bibitem{Capozzi:2017gqd}
F.~Capozzi, B.~Dasgupta, E.~Lisi, A.~Marrone, A.~Mirizzi, {Fast flavor
  conversions of supernova neutrinos: Classifying instabilities via dispersion
  relations}, Phys. Rev. D96~(4) (2017) 043016.
\newblock \href {http://arxiv.org/abs/1706.03360} {\path{arXiv:1706.03360}},
  \href {http://dx.doi.org/10.1103/PhysRevD.96.043016}
  {\path{doi:10.1103/PhysRevD.96.043016}}.

\bibitem{Richers:2019grc}
S.~A. Richers, G.~C. McLaughlin, J.~P. Kneller, A.~Vlasenko, {Neutrino Quantum
  Kinetics in Compact Objects}, Phys. Rev. D99~(12) (2019) 123014.
\newblock \href {http://arxiv.org/abs/1903.00022} {\path{arXiv:1903.00022}},
  \href {http://dx.doi.org/10.1103/PhysRevD.99.123014}
  {\path{doi:10.1103/PhysRevD.99.123014}}.

\bibitem{Dasgupta:2016dbv}
B.~Dasgupta, A.~Mirizzi, M.~Sen, {Fast neutrino flavor conversions near the
  supernova core with realistic flavor-dependent angular distributions}, JCAP
  1702~(02) (2017) 019.
\newblock \href {http://arxiv.org/abs/1609.00528} {\path{arXiv:1609.00528}},
  \href {http://dx.doi.org/10.1088/1475-7516/2017/02/019}
  {\path{doi:10.1088/1475-7516/2017/02/019}}.

\bibitem{Abbar:2017pkh}
S.~Abbar, H.~Duan, {Fast neutrino flavor conversion: roles of dense matter and
  spectrum crossing}, Phys. Rev. D98~(4) (2018) 043014.
\newblock \href {http://arxiv.org/abs/1712.07013} {\path{arXiv:1712.07013}},
  \href {http://dx.doi.org/10.1103/PhysRevD.98.043014}
  {\path{doi:10.1103/PhysRevD.98.043014}}.

\bibitem{Abbar:2018beu}
S.~Abbar, M.~C. Volpe, {On Fast Neutrino Flavor Conversion Modes in the
  Nonlinear Regime}, Phys. Lett. B790 (2019) 545--550.
\newblock \href {http://arxiv.org/abs/1811.04215} {\path{arXiv:1811.04215}},
  \href {http://dx.doi.org/10.1016/j.physletb.2019.02.002}
  {\path{doi:10.1016/j.physletb.2019.02.002}}.

\bibitem{Capozzi:2018clo}
F.~Capozzi, B.~Dasgupta, A.~Mirizzi, M.~Sen, G.~Sigl, {Collisional triggering
  of fast flavor conversions of supernova neutrinos}, Phys. Rev. Lett. 122~(9)
  (2019) 091101.
\newblock \href {http://arxiv.org/abs/1808.06618} {\path{arXiv:1808.06618}},
  \href {http://dx.doi.org/10.1103/PhysRevLett.122.091101}
  {\path{doi:10.1103/PhysRevLett.122.091101}}.

\bibitem{Martin:2019gxb}
J.~D. Martin, C.~Yi, H.~Duan, {Dynamic fast flavor oscillation waves in dense
  neutrino gases}\href {http://arxiv.org/abs/1909.05225}
  {\path{arXiv:1909.05225}}.

\bibitem{Capozzi:2019lso}
F.~Capozzi, G.~Raffelt, T.~Stirner, {Fast Neutrino Flavor Conversion:
  Collective Motion vs. Decoherence}, JCAP 1909 (2019) 002.
\newblock \href {http://arxiv.org/abs/1906.08794} {\path{arXiv:1906.08794}},
  \href {http://dx.doi.org/10.1088/1475-7516/2019/09/002}
  {\path{doi:10.1088/1475-7516/2019/09/002}}.

\bibitem{Doring:2019axc}
C.~Döring, R.~S.~L. Hansen, M.~Lindner, {Stability of three neutrino flavor
  conversion in supernovae}, JCAP 1908 (2019) 003.
\newblock \href {http://arxiv.org/abs/1905.03647} {\path{arXiv:1905.03647}},
  \href {http://dx.doi.org/10.1088/1475-7516/2019/08/003}
  {\path{doi:10.1088/1475-7516/2019/08/003}}.

\bibitem{Chakraborty:2019wxe}
M.~Chakraborty, S.~Chakraborty, {Three flavor neutrino conversions in
  supernovae: Slow $\&$ Fast instabilities}\href
  {http://arxiv.org/abs/1909.10420} {\path{arXiv:1909.10420}}.

\bibitem{Johns:2019izj}
L.~Johns, H.~Nagakura, G.~M. Fuller, A.~Burrows, {Neutrino oscillations in
  supernovae: angular moments and fast instabilities}\href
  {http://arxiv.org/abs/1910.05682} {\path{arXiv:1910.05682}}.

\bibitem{Cirigliano:2017hmk}
V.~Cirigliano, M.~W. Paris, S.~Shalgar, {Effect of collisions on neutrino
  flavor inhomogeneity in a dense neutrino gas}, Phys. Lett. B774 (2017)
  258--267.
\newblock \href {http://arxiv.org/abs/1706.07052} {\path{arXiv:1706.07052}},
  \href {http://dx.doi.org/10.1016/j.physletb.2017.09.039}
  {\path{doi:10.1016/j.physletb.2017.09.039}}.

\bibitem{Duan:2014gfa}
H.~Duan, S.~Shalgar, {Flavor instabilities in the neutrino line model}, Phys.
  Lett. B747 (2015) 139--143.
\newblock \href {http://arxiv.org/abs/1412.7097} {\path{arXiv:1412.7097}},
  \href {http://dx.doi.org/10.1016/j.physletb.2015.05.057}
  {\path{doi:10.1016/j.physletb.2015.05.057}}.

\bibitem{Chakraborty:2015tfa}
S.~Chakraborty, R.~S. Hansen, I.~Izaguirre, G.~Raffelt, {Self-induced flavor
  conversion of supernova neutrinos on small scales}, JCAP 1601~(01) (2016)
  028.
\newblock \href {http://arxiv.org/abs/1507.07569} {\path{arXiv:1507.07569}},
  \href {http://dx.doi.org/10.1088/1475-7516/2016/01/028}
  {\path{doi:10.1088/1475-7516/2016/01/028}}.

\bibitem{Abbar:2015mca}
S.~Abbar, H.~Duan, S.~Shalgar, {Flavor instabilities in the multiangle neutrino
  line model}, Phys. Rev. D92~(6) (2015) 065019.
\newblock \href {http://arxiv.org/abs/1507.08992} {\path{arXiv:1507.08992}},
  \href {http://dx.doi.org/10.1103/PhysRevD.92.065019}
  {\path{doi:10.1103/PhysRevD.92.065019}}.

\bibitem{Abbar:2015fwa}
S.~Abbar, H.~Duan, {Neutrino flavor instabilities in a time-dependent supernova
  model}, Phys. Lett. B751 (2015) 43--47.
\newblock \href {http://arxiv.org/abs/1509.01538} {\path{arXiv:1509.01538}},
  \href {http://dx.doi.org/10.1016/j.physletb.2015.10.019}
  {\path{doi:10.1016/j.physletb.2015.10.019}}.

\bibitem{Dasgupta:2015iia}
B.~Dasgupta, A.~Mirizzi, {Temporal Instability Enables Neutrino Flavor
  Conversions Deep Inside Supernovae}, Phys. Rev. D92~(12) (2015) 125030.
\newblock \href {http://arxiv.org/abs/1509.03171} {\path{arXiv:1509.03171}},
  \href {http://dx.doi.org/10.1103/PhysRevD.92.125030}
  {\path{doi:10.1103/PhysRevD.92.125030}}.

\bibitem{Hansen:2019iop}
R.~S.~L. Hansen, A.~Y. Smirnov, {Effect of extended neutrino production region
  on collective oscillations in supernovae}\href
  {http://arxiv.org/abs/1905.13670} {\path{arXiv:1905.13670}}.

\bibitem{Tamborra:2017ubu}
I.~Tamborra, L.~Huedepohl, G.~Raffelt, H.-T. Janka, {Flavor-dependent neutrino
  angular distribution in core-collapse supernovae}, Astrophys. J. 839 (2017)
  132.
\newblock \href {http://arxiv.org/abs/1702.00060} {\path{arXiv:1702.00060}},
  \href {http://dx.doi.org/10.3847/1538-4357/aa6a18}
  {\path{doi:10.3847/1538-4357/aa6a18}}.

\bibitem{Shalgar:2019kzy}
S.~Shalgar, I.~Tamborra, {On the Occurrence of Crossings Between the Angular
  Distributions of Electron Neutrinos and Antineutrinos in the Supernova Core},
  Astrophys. J. 883 (2019) 80.
\newblock \href {http://arxiv.org/abs/1904.07236} {\path{arXiv:1904.07236}},
  \href {http://dx.doi.org/10.3847/1538-4357/ab38ba}
  {\path{doi:10.3847/1538-4357/ab38ba}}.

\bibitem{Morinaga:2019wsv}
T.~Morinaga, H.~Nagakura, C.~Kato, S.~Yamada, {A new possibility of the fast
  neutrino-flavor conversion in the pre-shock region of core-collapse
  supernova}\href {http://arxiv.org/abs/1909.13131} {\path{arXiv:1909.13131}}.

\bibitem{Tamborra:2014aua}
I.~Tamborra, F.~Hanke, H.-T. Janka, B.~M{\"u}ller, G.~G. Raffelt, A.~Marek,
  {Self-sustained asymmetry of lepton-number emission: A new phenomenon during
  the supernova shock-accretion phase in three dimensions}, Astrophys. J.
  792~(2) (2014) 96.
\newblock \href {http://arxiv.org/abs/1402.5418} {\path{arXiv:1402.5418}},
  \href {http://dx.doi.org/10.1088/0004-637X/792/2/96}
  {\path{doi:10.1088/0004-637X/792/2/96}}.

\bibitem{Vartanyan:2019ssu}
D.~Vartanyan, A.~Burrows, D.~Radice, {Temporal and Angular Variations of 3D
  Core-Collapse Supernova Emissions and their Physical Correlations}\href
  {http://arxiv.org/abs/1906.08787} {\path{arXiv:1906.08787}}, \href
  {http://dx.doi.org/10.1093/mnras/stz2307} {\path{doi:10.1093/mnras/stz2307}}.

\bibitem{Sugiura:2019xuv}
K.~Sugiura, K.~Takahashi, S.~Yamada, {Linear Analysis of the Shock Instability
  in Core-collapse Supernovae: Influences of Acoustic Power and Fluctuations of
  Neutrino Luminosity}, Astrophys. J. 874~(1) (2019) 28.
\newblock \href {http://arxiv.org/abs/1903.00480} {\path{arXiv:1903.00480}},
  \href {http://dx.doi.org/10.3847/1538-4357/ab08a2}
  {\path{doi:10.3847/1538-4357/ab08a2}}.

\bibitem{Walk:2019ier}
L.~Walk, I.~Tamborra, H.-T. Janka, A.~Summa, {Effects of the standing
  accretion-shock instability and the lepton-emission self-sustained asymmetry
  in the neutrino emission of rotating supernovae}, Phys. Rev. D100 (2019)
  063018.
\newblock \href {http://arxiv.org/abs/1901.06235} {\path{arXiv:1901.06235}}.

\bibitem{Glas:2018vcs}
R.~Glas, H.~T. Janka, T.~Melson, G.~Stockinger, O.~Just, {Effects of LESA in
  Three-Dimensional Supernova Simulations with Multi-Dimensional and
  Ray-by-Ray-plus Neutrino Transport}\href {http://arxiv.org/abs/1809.10150}
  {\path{arXiv:1809.10150}}, \href {http://dx.doi.org/10.3847/1538-4357/ab275c}
  {\path{doi:10.3847/1538-4357/ab275c}}.

\bibitem{OConnor:2018tuw}
E.~P. O'Connor, S.~M. Couch, {Exploring Fundamentally Three-dimensional
  Phenomena in High-fidelity Simulations of Core-collapse Supernovae},
  Astrophys. J. 865~(2) (2018) 81.
\newblock \href {http://arxiv.org/abs/1807.07579} {\path{arXiv:1807.07579}},
  \href {http://dx.doi.org/10.3847/1538-4357/aadcf7}
  {\path{doi:10.3847/1538-4357/aadcf7}}.

\bibitem{Janka:2016fox}
H.~T. Janka, T.~Melson, A.~Summa, {Physics of Core-Collapse Supernovae in Three
  Dimensions: a Sneak Preview}, Ann. Rev. Nucl. Part. Sci. 66 (2016) 341--375.
\newblock \href {http://arxiv.org/abs/1602.05576} {\path{arXiv:1602.05576}},
  \href {http://dx.doi.org/10.1146/annurev-nucl-102115-044747}
  {\path{doi:10.1146/annurev-nucl-102115-044747}}.

\bibitem{Tamborra:2014hga}
I.~Tamborra, G.~Raffelt, F.~Hanke, H.-T. Janka, B.~Mueller, {Neutrino emission
  characteristics and detection opportunities based on three-dimensional
  supernova simulations}, Phys. Rev. D90~(4) (2014) 045032.
\newblock \href {http://arxiv.org/abs/1406.0006} {\path{arXiv:1406.0006}},
  \href {http://dx.doi.org/10.1103/PhysRevD.90.045032}
  {\path{doi:10.1103/PhysRevD.90.045032}}.

\bibitem{Nagakura:2019evv}
H.~Nagakura, K.~Sumiyoshi, S.~Yamada, {Possible early linear acceleration of
  proto-neutron stars via asymmetric neutrino emission in core-collapse
  supernovae}, Astrophys. J. 880~(2) (2019) L28.
\newblock \href {http://arxiv.org/abs/1907.04863} {\path{arXiv:1907.04863}},
  \href {http://dx.doi.org/10.3847/2041-8213/ab30ca}
  {\path{doi:10.3847/2041-8213/ab30ca}}.

\bibitem{Abbar:2018shq}
S.~Abbar, H.~Duan, K.~Sumiyoshi, T.~Takiwaki, M.~C. Volpe, {On the occurrence
  of fast neutrino flavor conversions in multidimensional supernova models},
  Phys. Rev. D100~(4) (2019) 043004.
\newblock \href {http://arxiv.org/abs/1812.06883} {\path{arXiv:1812.06883}},
  \href {http://dx.doi.org/10.1103/PhysRevD.100.043004}
  {\path{doi:10.1103/PhysRevD.100.043004}}.

\bibitem{Sumiyoshi:2012za}
K.~Sumiyoshi, S.~Yamada, {Neutrino Transfer in Three Dimensions for
  Core-Collapse Supernovae. I. Static Configurations}, Astrophys. J. Suppl. 199
  (2012) 17.
\newblock \href {http://arxiv.org/abs/1201.2244} {\path{arXiv:1201.2244}},
  \href {http://dx.doi.org/10.1088/0067-0049/199/1/17}
  {\path{doi:10.1088/0067-0049/199/1/17}}.

\bibitem{Sumiyoshi:2014qua}
K.~Sumiyoshi, T.~Takiwaki, H.~Matsufuru, S.~Yamada, {Multi-dimensional Features
  of Neutrino Transfer in Core-Collapse Supernovae}, Astrophys. J. Suppl. 216
  (2015) 5.
\newblock \href {http://arxiv.org/abs/1403.4476} {\path{arXiv:1403.4476}},
  \href {http://dx.doi.org/10.1088/0067-0049/216/1/5}
  {\path{doi:10.1088/0067-0049/216/1/5}}.

\bibitem{Nagakura:2017mnp}
H.~Nagakura, W.~Iwakami, S.~Furusawa, H.~Okawa, A.~Harada, K.~Sumiyoshi,
  S.~Yamada, H.~Matsufuru, A.~Imakura, {Simulations of core-collapse supernovae
  in spatial axisymmetry with full Boltzmann neutrino transport}, Astrophys. J.
  854~(2) (2018) 136.
\newblock \href {http://arxiv.org/abs/1702.01752} {\path{arXiv:1702.01752}},
  \href {http://dx.doi.org/10.3847/1538-4357/aaac29}
  {\path{doi:10.3847/1538-4357/aaac29}}.

\bibitem{Azari:2019jvr}
M.~Delfan~Azari, S.~Yamada, T.~Morinaga, W.~Iwakami, H.~Okawa, H.~Nagakura,
  K.~Sumiyoshi, {Linear Analysis of Fast-Pairwise Collective Neutrino
  Oscillations in Core-Collapse Supernovae based on the Results of Boltzmann
  Simulations}, Phys. Rev. D99~(10) (2019) 103011.
\newblock \href {http://arxiv.org/abs/1902.07467} {\path{arXiv:1902.07467}},
  \href {http://dx.doi.org/10.1103/PhysRevD.99.103011}
  {\path{doi:10.1103/PhysRevD.99.103011}}.

\bibitem{Nagakura:2019sig}
H.~Nagakura, T.~Morinaga, C.~Kato, S.~Yamada, {Fast-pairwise collective
  neutrino oscillations associated with asymmetric neutrino emissions in
  core-collapse supernova}\href {http://arxiv.org/abs/1910.04288}
  {\path{arXiv:1910.04288}}.

\bibitem{DelfanAzari:2019tez}
M.~Delfan~Azari, S.~Yamada, T.~Morinaga, H.~Nagakura, S.~Furusawa, A.~Harada,
  H.~Okawa, W.~Iwakami, K.~Sumiyoshi, {Fast collective neutrino oscillations
  inside the neutrino sphere in core-collapse supernovae}\href
  {http://arxiv.org/abs/1910.06176} {\path{arXiv:1910.06176}}.

\bibitem{Sigl:1992fn}
G.~Sigl, G.~Raffelt, {General kinetic description of relativistic mixed
  neutrinos}, Nucl. Phys. B406 (1993) 423--451.
\newblock \href {http://dx.doi.org/10.1016/0550-3213(93)90175-O}
  {\path{doi:10.1016/0550-3213(93)90175-O}}.

\bibitem{Strack:2005ux}
P.~Strack, A.~Burrows, A generalized boltzmann formalism for oscillating
  neutrinos, Phys. Rev. D71 (2005) 093004.
\newblock \href {http://arxiv.org/abs/hep-ph/0504035}
  {\path{arXiv:hep-ph/0504035}}.

\bibitem{Cardall:2007zw}
C.~Y. Cardall, {Liouville equations for neutrino distribution matrices}, Phys.
  Rev. D78 (2008) 085017.
\newblock \href {http://arxiv.org/abs/0712.1188} {\path{arXiv:0712.1188}},
  \href {http://dx.doi.org/10.1103/PhysRevD.78.085017}
  {\path{doi:10.1103/PhysRevD.78.085017}}.

\bibitem{Volpe:2013jgr}
C.~Volpe, D.~Väänänen, C.~Espinoza, {Extended evolution equations for
  neutrino propagation in astrophysical and cosmological environments}, Phys.
  Rev. D87~(11) (2013) 113010.
\newblock \href {http://arxiv.org/abs/1302.2374} {\path{arXiv:1302.2374}},
  \href {http://dx.doi.org/10.1103/PhysRevD.87.113010}
  {\path{doi:10.1103/PhysRevD.87.113010}}.

\bibitem{Vlasenko:2013fja}
A.~Vlasenko, G.~M. Fuller, V.~Cirigliano, {Neutrino Quantum Kinetics}, Phys.
  Rev. D89~(10) (2014) 105004.
\newblock \href {http://arxiv.org/abs/1309.2628} {\path{arXiv:1309.2628}},
  \href {http://dx.doi.org/10.1103/PhysRevD.89.105004}
  {\path{doi:10.1103/PhysRevD.89.105004}}.

\bibitem{Wolfenstein:1977ue}
L.~Wolfenstein, {Neutrino Oscillations in Matter}, Phys. Rev. D17 (1978)
  2369--2374, [,294(1977)].
\newblock \href {http://dx.doi.org/10.1103/PhysRevD.17.2369}
  {\path{doi:10.1103/PhysRevD.17.2369}}.

\bibitem{Mikheev:1986gs}
S.~P. Mikheyev, A.~{\relax Yu}. Smirnov, {Resonance Amplification of
  Oscillations in Matter and Spectroscopy of Solar Neutrinos}, Sov. J. Nucl.
  Phys. 42 (1985) 913--917, [,305(1986)].

\bibitem{Fuller:1987aa}
G.~M. Fuller, R.~W. Mayle, J.~R. Wilson, D.~N. Schramm, Resonant neutrino
  oscillations and stellar collapse, Astrophys. J. 322 (1987) 795.

\bibitem{Notzold:1988kx}
D.~N\"{o}tzold, G.~Raffelt, Neutrono dispersion at finite temperature and
  density, Nucl. Phys. B307 (1988) 924.

\bibitem{Pantaleone:1992xh}
J.~T. Pantaleone, {Dirac neutrinos in dense matter}, Phys. Rev. D46 (1992)
  510--523.
\newblock \href {http://dx.doi.org/10.1103/PhysRevD.46.510}
  {\path{doi:10.1103/PhysRevD.46.510}}.

\bibitem{Banerjee:2011fj}
A.~Banerjee, A.~Dighe, G.~Raffelt, {Linearized flavor-stability analysis of
  dense neutrino streams}, Phys. Rev. D84 (2011) 053013.
\newblock \href {http://arxiv.org/abs/1107.2308} {\path{arXiv:1107.2308}},
  \href {http://dx.doi.org/10.1103/PhysRevD.84.053013}
  {\path{doi:10.1103/PhysRevD.84.053013}}.

\bibitem{Vaananen:2013qja}
D.~Väänänen, C.~Volpe, {Linearizing neutrino evolution equations including
  neutrino-antineutrino pairing correlations}, Phys. Rev. D88 (2013) 065003.
\newblock \href {http://arxiv.org/abs/1306.6372} {\path{arXiv:1306.6372}},
  \href {http://dx.doi.org/10.1103/PhysRevD.88.065003}
  {\path{doi:10.1103/PhysRevD.88.065003}}.

\bibitem{sturrock1958kinematics}
P.~A. Sturrock, Kinematics of growing waves, Physical Review 112~(5) (1958)
  1488.

\bibitem{Yi:2019hrp}
C.~Yi, L.~Ma, J.~D. Martin, H.~Duan, {Dispersion relation of the fast neutrino
  oscillation wave}, Phys. Rev. D99~(6) (2019) 063005.
\newblock \href {http://arxiv.org/abs/1901.01546} {\path{arXiv:1901.01546}},
  \href {http://dx.doi.org/10.1103/PhysRevD.99.063005}
  {\path{doi:10.1103/PhysRevD.99.063005}}.

\bibitem{Bruenn:1985en}
S.~W. Bruenn, {Stellar core collapse: Numerical model and infall epoch},
  Astrophys. J. Suppl. 58 (1985) 771--841.
\newblock \href {http://dx.doi.org/10.1086/191056} {\path{doi:10.1086/191056}}.

\bibitem{Lattimer:1991nc}
J.~M. Lattimer, F.~D. Swesty, {A Generalized equation of state for hot, dense
  matter}, Nucl. Phys. A535 (1991) 331--376.
\newblock \href {http://dx.doi.org/10.1016/0375-9474(91)90452-C}
  {\path{doi:10.1016/0375-9474(91)90452-C}}.

\bibitem{Takiwaki:2011db}
T.~Takiwaki, K.~Kotake, Y.~Suwa, {Three-dimensional Hydrodynamic Core-Collapse
  Supernova Simulations for an $11.2 M_{\odot}$ Star with Spectral Neutrino
  Transport}, Astrophys. J. 749 (2012) 98.
\newblock \href {http://arxiv.org/abs/1108.3989} {\path{arXiv:1108.3989}},
  \href {http://dx.doi.org/10.1088/0004-637X/749/2/98}
  {\path{doi:10.1088/0004-637X/749/2/98}}.

\bibitem{Takiwaki:2013cqa}
T.~Takiwaki, K.~Kotake, Y.~Suwa, {A Comparison of Two- and Three-dimensional
  Neutrino-hydrodynamics simulations of Core-collapse Supernovae}, Astrophys.
  J. 786 (2014) 83.
\newblock \href {http://arxiv.org/abs/1308.5755} {\path{arXiv:1308.5755}},
  \href {http://dx.doi.org/10.1088/0004-637X/786/2/83}
  {\path{doi:10.1088/0004-637X/786/2/83}}.

\bibitem{Harada:2018ubo}
A.~Harada, H.~Nagakura, W.~Iwakami, H.~Okawa, S.~Furusawa, H.~Matsufuru,
  K.~Sumiyoshi, S.~Yamada, {On the Neutrino Distributions in Phase Space for
  the Rotating Core-Collapse Supernova Simulated with a
  Boltzmann-Neutrino-Radiation-Hydrodynamics Code}, Astrophys. J. 872~(2)
  (2019) 181.
\newblock \href {http://arxiv.org/abs/1810.12316} {\path{arXiv:1810.12316}},
  \href {http://dx.doi.org/10.3847/1538-4357/ab0203}
  {\path{doi:10.3847/1538-4357/ab0203}}.

\bibitem{Duan:2007fw}
H.~Duan, G.~M. Fuller, Y.-Z. Qian, {A Simple Picture for Neutrino Flavor
  Transformation in Supernovae}, Phys. Rev. D76 (2007) 085013.
\newblock \href {http://arxiv.org/abs/0706.4293} {\path{arXiv:0706.4293}},
  \href {http://dx.doi.org/10.1103/PhysRevD.76.085013}
  {\path{doi:10.1103/PhysRevD.76.085013}}.

\bibitem{mangano:2014zda}
G.~Mangano, A.~Mirizzi, N.~Saviano, {Damping the neutrino flavor pendulum by
  breaking homogeneity}, Phys. Rev. D89~(7) (2014) 073017.
\newblock \href {http://arxiv.org/abs/1403.1892} {\path{arXiv:1403.1892}},
  \href {http://dx.doi.org/10.1103/PhysRevD.89.073017}
  {\path{doi:10.1103/PhysRevD.89.073017}}.

\bibitem{Furusawa:2011wh}
S.~Furusawa, S.~Yamada, K.~Sumiyoshi, H.~Suzuki, {A new baryonic equation of
  state at sub-nuclear densities for core-collapse simulations}, Astrophys. J.
  738 (2011) 178.
\newblock \href {http://arxiv.org/abs/1103.6129} {\path{arXiv:1103.6129}},
  \href {http://dx.doi.org/10.1088/0004-637X/738/2/178}
  {\path{doi:10.1088/0004-637X/738/2/178}}.

\end{thebibliography}

\clearpage
\appendix

\end{document}